\documentclass[prd,notitlepage,showpacs,nofootinbib,superscriptaddress]{revtex4-2}

\usepackage[utf8]{inputenc} 

\usepackage{amsmath}
\usepackage{amssymb}
\usepackage{float}
\usepackage{comment}
\usepackage{slashed}
\usepackage[normalem]{ulem}
\usepackage{bbold}
\usepackage{cancel}
\usepackage{graphicx}
\usepackage{booktabs}
\usepackage{multirow}
\usepackage{rotating}
\usepackage{caption}
\usepackage{subcaption}
\usepackage{tabulary}

\newcolumntype{K}[1]{>{\centering\arraybackslash}m{#1}}
\def\gsim{\raise0.3ex\hbox{$\;>$\kern-0.75em\raise-1.1ex\hbox{$\sim\;$}}}
\def\lsim{\raise0.3ex\hbox{$\;<$\kern-0.75em\raise-1.1ex\hbox{$\sim\;$}}}

\usepackage[usenames,dvipsnames]{color}

\newcommand {\ignore}[1]{}

\usepackage[colorinlistoftodos]{todonotes}
\usepackage[colorlinks=true,linkcolor=darkblue,citecolor=darkblue,urlcolor=darkblue, pdfborder={0 0 0}]{hyperref}
\usepackage[normalem]{ulem}

\usepackage{caption}
\usepackage{subcaption}
\definecolor{linkcolor}{rgb}{0,0,0.8}

\definecolor{darkgreen}{rgb}{0,0.5,0}
\definecolor{darkred}{rgb}{0.6,0,0}
\definecolor{brown}{rgb}{0.59, 0.29, 0.0}
\definecolor{mightnightblue}{RGB}{25,25,112}
\definecolor{darkblue}{rgb}{0,0,0.8}

\newcommand {\darkblue} {\color{darkblue}}

\newcommand{\cbeta}{c_\beta}
\newcommand{\sbeta}{s_\beta}
\newcommand{\calpha}{c_\alpha}
\newcommand{\salpha}{s_\alpha}

\def\SM{$\mathrm{SU(3)_c \otimes SU(2)_L \otimes U(1)_Y}$ }

\usepackage[force]{feynmp-auto}

\begin{document}
\vspace{-3cm}
\begin{flushright}
IFT-UAM-CSIC-23-86
\end{flushright}

\title{\darkblue \Large Confronting the 95 GeV excesses within the UN2HDM}

\author{\bf J.~A.~Aguilar-Saavedra}
\email{ja.a.s@csic.es}
\affiliation{Instituto de F\'{\i}sica Te\'{o}rica UAM-CSIC, Campus de Cantoblanco, E-28049 Madrid, Spain}
\author{\bf H.~B. C\^amara}\email{henrique.b.camara@tecnico.ulisboa.pt}
\affiliation{Departamento de F\'{\i}sica and CFTP, Instituto Superior T\'ecnico, Universidade de Lisboa, Av. Rovisco Pais 1, 1049-001 Lisboa, Portugal}
\author{\bf F. R. Joaquim}\email{filipe.joaquim@tecnico.ulisboa.pt}
\affiliation{Departamento de F\'{\i}sica and CFTP, Instituto Superior T\'ecnico, Universidade de Lisboa, Av. Rovisco Pais 1, 1049-001 Lisboa, Portugal}
\author{\bf J. F. Seabra}
\email{joao.f.seabra@tecnico.ulisboa.pt}
\affiliation{Instituto de F\'{\i}sica Te\'{o}rica UAM-CSIC, Campus de Cantoblanco, E-28049 Madrid, Spain}
\affiliation{Departamento de F\'{\i}sica and CFTP, Instituto Superior T\'ecnico, Universidade de Lisboa, Av. Rovisco Pais 1, 1049-001 Lisboa, Portugal}

%%%%%%%%%%%%%%%%%%%%%%%%%%%%%%%%%%%%%%%%%%%%%%%%%%%%%%%%%%%%%%%%%%%%%%%%%%%%%
\begin{abstract}
\vspace{0.2cm}
\begin{center}
{ \center  \bf ABSTRACT}\\    
\end{center}
We consider the small excesses around 95 GeV found in several searches for a new scalar in $\gamma \gamma$, $\tau \tau$ and $b \bar b$ final states. Instead of trying to accommodate them all, as is usually done in the literature, in the context of a given Standard Model~(SM) extension, we investigate whether it would be possible that one or two of these excesses correspond to an actual new scalar, while the remaining ones are merely statistical fluctuations. To this end, we use as benchmark model the UN2HDM, a SM extension with one scalar doublet, one scalar singlet, and an extra $\text{U}(1)'$ symmetry, which has been previously studied in the context of multiboson cascade decays. We show that most of the possibilities where the excesses in one or two of these channels disappear in the future can be accommodated by type-I or type-III UN2HDMs.
\end{abstract}
%%%%%%%%%%%%%%%%%%%%%%%%%%%%%%%%%%%%%%%%%%%%%%%%%%%%%%%%%%%%%%%%%%%%%%%%%%%%%

\maketitle
\noindent

%%%%%%%%%%%%%%%%%%%%%%%%%%%%%%%%%%%%%%%%%%%%%%%%%%%%%%%%%%%%%%%%%%%%%%%%%%%%%
\section{Introduction}
%%%%%%%%%%%%%%%%%%%%%%%%%%%%%%%%%%%%%%%%%%%%%%%%%%%%%%%%%%%%%%%%%%%%%%%%%%%%%

In the last decade, the ATLAS and CMS experiments have been collecting data from $pp$ collisions at the Large Hadron Collider (LHC) with the purpose of measuring particle properties and searching for physics beyond the Standard Model (SM). Although the experimental results seem to be following the SM score, in some cases deviations with respect to what is expected appear. Typically, if such anomalies reach a statistical significance close to $3\sigma$, they trigger the attention of the theoretical community towards explaining them in the context of SM extensions. This has been the case for excess events reported by experiments around a mass of $95$~GeV. Namely,
\begin{itemize}
\item In 2018, the CMS Collaboration reported a $2.8\sigma$ local excess in the diphoton invariant-mass distribution at $m_{\gamma \gamma} = 95$ GeV, combining 19.7 fb$^{-1}$ of LHC data at 8 TeV and 35.9 fb$^{-1}$ at 13 TeV~\cite{CMS:2018cyk}. By that time, the ATLAS collaboration did not find a significant excess at this mass with 80 fb$^{-1}$~\cite{ATLAS:2018xad}. However, their sensitivity was smaller and, thus, their limits were not in tension with the CMS anomaly. A recent CMS analysis of the full Run 2 dataset with 138 fb$^{-1}$~\cite{CMS:2023yay}, fell short of confirming the excess, with the local significance maintained slightly below the $3\sigma$ level and, more importantly, without a significant excess appearing in 2017-2018 data. 
\item Recently, ATLAS released the results on their searches for diphoton resonances in the 66-110 GeV mass range using 140~fb$^{-1}$ of 13 TeV $pp$ collisions (full Run 2 LHC dataset)~\cite{ATLAS-CONF-2023-035}. With respect to their previous analysis~\cite{ATLAS:2018xad}, the use of multivariate analysis techniques in background mitigation and event classification, resulted in a improvement of the sensitivity to physics beyond the SM. The new ATLAS analysis reveals an excess in the diphoton channel at an invariant mass around 95 GeV, with a local significance of $1.7\sigma$, which is too small to merit attention on its own but happens at the same mass of the previous one.
\item In a dedicated search for additional Higgs bosons $\phi$ and vector leptoquarks in $\tau\tau$ final states using the full LHC Run 2 dataset~\cite{CMS:2022goy}, CMS found a $3.1\sigma$ excess of events for $b\bar{b},gg \rightarrow \phi \rightarrow \tau\tau$ at an invariant mass $m_{\tau \tau}\simeq 100$ GeV. Given the poor mass resolution in this final state when compared to the $\gamma\gamma$ channel, this local excess seems to be compatible with the diphoton one.
\item In relation to these hints, the Large Electron Positron (LEP) experiments ALEPH, DELPHI, L3, OPAL released in 2006 the results of a statistical combination of their data, showing an excess in the $e^+e^- \rightarrow Z\phi\,(\phi\rightarrow b\bar{b})$ mode at $m_{b\bar{b}}\simeq 98$~GeV with a local significance of $2.3\sigma$~\cite{LEPWorkingGroupforHiggsbosonsearches:2003ing,Azatov:2012bz,Cao:2016uwt}. Bearing in mind the limited mass resolution for dijets at LEP, the $b\bar{b}$ excess could originate from the same particle responsible for the $\gamma\gamma$ and $\tau\tau$ anomalies summarized above.
\end{itemize}
The size of the above excesses can be conveniently expressed in terms of a signal-strength parameter $\mu$, which relates the cross section of the potential signal generating the event excess with the production cross section of a SM-like scalar with the mass at which the excess is located. For the CMS and ATLAS diphoton excesses $\mu_{\gamma \gamma}$ is~\cite{Biekotter:2023oen}
\begin{align}
\mu_{\gamma \gamma}^\text{CMS} = 0.33^{+0.19}_{-0.12}\;,\; \mu_{\gamma \gamma}^\text{ATLAS} = 0.21\pm 0.12
\;,\; \mu_{\gamma \gamma}^\text{exp} = \mu_{\gamma \gamma}^\text{ATLAS+CMS} & = 0.27^{+0.10}_{-0.09}\;,
\label{eq:mugg}
\end{align}
where $\mu_{\gamma \gamma}^\text{exp}$ has been obtained combining $\mu_{\gamma \gamma}^\text{CMS}$ and $\mu_{\gamma \gamma}^\text{ATLAS}$ in quadrature, while for the CMS $\tau\tau$ and the LEP $b\bar{b}$ ones
\begin{align}
\mu_{\tau \tau}^\text{exp} = 1.2 \pm 0.5\;,\;
\mu_{b \bar b}^\text{exp} = 0.117 \pm 0.057\,.
\label{eq:muttbb}
\end{align}

\noindent The appearance of the aforementioned anomalies triggered several studies on the possibility of accommodating them in the context of BSM models. For our purposes we highlight the two Higgs-doublet model (2HDM)~\cite{Fox:2017uwr, Haisch:2017gql, Azevedo:2023zkg,Belyaev:2023xnv}, and also its extensions featuring an additional real~\cite{Biekotter:2019kde, Biekotter:2021qbc, Heinemeyer:2021msz, Biekotter:2022jyr} or complex~\cite{Heinemeyer:2021msz, Biekotter:2021ovi, Biekotter:2023jld, Biekotter:2023oen} scalar singlet (the so-called N2HDM models).
In the $\mathcal{Z}_2$-symmetric CP-conserving 2HDM, the type-I  Yukawa sector can accommodate $\gamma \gamma$ alone~\cite{Fox:2017uwr, Haisch:2017gql}, or together with $\tau \tau$~\cite{Azevedo:2023zkg}, the latter in tension with flavor observables. For the lepton-specific version diphoton and ditau cannot be accommodated simultaneously. The CP-violating type-I case can potentially explain all three excesses, albeit in conflict with electric dipole moment bounds~\cite{Azevedo:2023zkg}. In the general 2HDM, taking all relevant theoretical and experimental constraints, it was shown that, $\gamma\gamma$, $\tau\tau$ and $b\bar{b}$ excesses, can be explained simultaneously~\cite{Belyaev:2023xnv}.
Adding a real/complex singlet to the type II $\mathcal{Z}_2$-symmetric 2HDM, the $\gamma \gamma$ and $b \overline{b}$ excesses can be simultaneously accommodated, while in type-IV this happens for $\gamma \gamma$ and $b \overline{b}$ or $\gamma \gamma$ and $\tau \tau$~\cite{Biekotter:2023oen}.
Furthermore, it is also relevant to discuss scenarios featuring scalar singlet and vector-like fermions. These were only studied in the context of accommodating diphoton excess alone~\cite{Fox:2017uwr} and diphoton and dibottom simultaneously~\cite{Aguilar-Saavedra:2020wrj}.
Other extensions of the SM including, for instance, an $\mathrm{SU(2)_L}$ scalar triplet~\cite{Ashanujjaman:2023etj}, extended gauge sectors~\cite{Banik:2023ecr} and supersymmetry~\cite{Cao:2016uwt, Biekotter:2017xmf, Domingo:2018uim, Cao:2019ofo, Li:2022etb} have also been explored, as well as possible connections of the observed excesses to extra dimensions~\cite{Richard:2017kot}, $B$-anomalies~\cite{Liu:2018xsw}, dark matter~\cite{Cline:2019okt} and neutrino mass generation mechanisms~\cite{Escribano:2023hxj}.

Another example of what could be a deviation with respect to the SM was a $3.4\sigma$ bump near 2 TeV in an ATLAS search for hadronically-decaying diboson resonances with Run 1 data~\cite{ATLAS:2015xom}. This could stem from multiboson production originated in the cascade decay of a new resonance~\cite{Aguilar-Saavedra:2015rna, Aguilar-Saavedra:2015iew}. The persistence of such anomaly in ATLAS~\cite{ATLAS:2017zuf} and CMS~\cite{CMS:2017fgc} diboson searches with Run 2 data, with a small local significance of $2\sigma$, motivated the proposal of a `stealth boson' signature~\cite{Aguilar-Saavedra:2017zuc} with a boosted particle $S$ undergoing a cascade decay  $S\rightarrow A_1 A_2\rightarrow q\overline{q}q\overline{q}$, where $A_{1,2}$ can the be SM gauge bosons $W$ and $Z$, the SM Higgs boson, or new scalars. Recently, a bump near 2 TeV also appeared in the results of a CMS search for hadronically-decaying diboson resonances with the full Run 2 dataset, with a local significance of $3.6\sigma$~\cite{arXiv:2210.00043}. Minimal stealth boson models (MSBMs) where the heavy resonance $R$ is a color-singlet neutral gauge boson $Z'$ were proposed in~\cite{Aguilar-Saavedra:2019adu} and further explored in~\cite{Aguilar-Saavedra:2020wrj}. In these scenarios, the breaking of the new ${\rm U(1)}'$ gauge symmetry is ensured by two complex scalar singlets which also provide the aforementioned cascade decays. More recently, the same problem was tackled in the context of the N2HDM where a complex singlet breaks the ${\rm U(1)}'$ gauge symmetry~\cite{Aguilar-Saavedra:2022rvy} -- the UN2HDM model.

In this work we confront the $\gamma\gamma$, $\tau\tau$ and $b\bar{b}$ excesses described above in the UN2HDM, by investigating whether the results in Eqs.~\eqref{eq:mugg} and \eqref{eq:muttbb} can be accommodated in that model. Although we perform a $\chi^2$ analysis where all the excesses are simultaneously considered, we also explore the possibility that one or more of them (if not all!) might well be statistical fluctuations. This exercise is quite relevant since, in general, a new scalar will produce signals in all the three final states (as well as other less sensitive ones). In the likely case that one or more of the anomalies are washed out with more data, one will have to confront the viability of the remaining one(s) in the context of models with extended scalar sector. The paper is organized as follows. In Sec.~\ref{sec:model}, we present the type-I and III fermion sector versions of the UN2HDM, as well as its scalar potential. We describe, in Sec.~\ref{sec:numerical}, our numerical procedure and the constraints we impose on the model's parameter space. By the end of that section we also show our results and leave discussion for Sec.~\ref{sec:discussion}. Details regarding scalar parameter reconstruction, interactions in the mass-eigenstate basis and one-loop expression of the Higgs to diphoton process, are found in the appendices.

%%%%%%%%%%%%%%%%%%%%%%%%%%%%%%%%%%%%%%%%%%%%%%%%%%%%%%%%%%%%%%%%%%%%%%%%%%%%%
\section{The UN2HDM}
\label{sec:model}
%%%%%%%%%%%%%%%%%%%%%%%%%%%%%%%%%%%%%%%%%%%%%%%%%%%%%%%%%%%%%%%%%%%%%%%%%%%%%

The UN2HDM extends the SM gauge group $G_{\rm SM}$=\SM  with an extra U$(1)^\prime$ symmetry, featuring an additional neutral gauge boson $Z^\prime$. Its scalar content comprises two Higgs doublets~$\Phi_{1,2}$ and a complex scalar singlet~$\chi$. The latter, being charged under U$(1)^\prime$, renders the $Z^\prime$ massive via the Higgs mechanism. Due to strong experimental limits set by $Z^\prime$ leptonic decay searches, we consider the case of leptophobic $Z^\prime$. This requires the SM leptons and one of the Higgs doublets $\Phi_{1,2}$ to be uncharged under U$(1)^\prime$. The UN2HDM has been studied in the context of multiboson cascade decays when one of the Higgs doublets has U$(1)^\prime$ charge~\cite{Aguilar-Saavedra:2022rvy}. Furthermore, the SM quarks are charged under the new gauge group allowing for the new vector boson to be produced at hadron colliders. Overall, this implies additional fermionic content to cancel gauge anomalies, with the best candidates being charged and neutral vector-like leptons~(VLLs) which can be SU$(2)_L$ doublets and/or singlets~\footnote{These are vector-like only under the SM gauge group, with their right and left-handed components being charged differently under U$(1)^\prime$ (see Table~\ref{tab:TypeI} and~\ref{tab:TypeIII}).}. Note that, models featuring vector-like quarks~(VLQs) have also been proposed in the literature~\cite{Aguilar-Saavedra:2020wrj}. However, VLLs are more appealing since, if stable, the neutral ones can be viable fermionic dark matter~(DM) candidates. Thus, in this work we focus on the UN2HDM with VLLs.

As for the Yukawa sector, the 2HDM provides four different setups (type I-IV)~\cite{Branco:2011iw}. Namely, in type-I one Higgs doublet couples to all SM fermions. This scenario was previously considered in Ref.~\cite{Aguilar-Saavedra:2022rvy}. Here we also present a type-III (lepton-specific) version of the UN2HDM where one doublet couples to quarks and the other to the leptons. In the type-II (type-IV/flipped) version, one doublet couples to down-type (up-type) quarks and leptons while the other couples to up-type (down-type) quarks. These latter cases were recently studied in the S2HDM in light of the 95 GeV Higgs boson excess~\cite{Biekotter:2023jld}. In the UN2HDM framework, type-II and IV Yukawa sectors imply additional VLQs for gauge anomaly cancellation since the up and down quarks need to be charged differently under the U$(1)^\prime$. Henceforth, we will study the type-I and III UN2HDM models in regards to the observed excesses for a potential 95 GeV Higgs boson reported by LEP in the $b \overline{b}$ channel, CMS in the $\tau \tau$ and CMS and ATLAS for the $\gamma \gamma$ channels. In the following we present the aspects of the UN2HDM useful for our purposes, namely the fermion sector and scalar potential. The analysis of the gauge sector and further details on the scalar interactions can be found in Ref.~\cite{Aguilar-Saavedra:2022rvy}.

\subsection{Type-I/III fermion sectors}
\label{sec:TypeI}

\begin{table*}[t!]
\setlength{\tabcolsep}{-1pt}
\centering
\setlength{\tabcolsep}{20pt}
\begin{tabular}{|c|c|c|c|c|}
		\hline
 &Fields& $G_{\rm SM}$ &    U($1$)$^\prime$ \\ 
		\hline \hline
		\multirow{5}{*}{SM fermions}
&$q_L$&($\mathbf{3},\mathbf{2}, 1/6$)& $Y^\prime$  \\
&$u_R$&($\mathbf{3},\mathbf{1}, 2/3$)&  $Y^\prime$   \\
&$d_R$&($\mathbf{3},\mathbf{1}, -1/3$)&  $Y^\prime$   \\
&$\ell_L$&($\mathbf{1},\mathbf{2}, -1/2$)&  $0$   \\
&$e_R$&($\mathbf{1},\mathbf{1}, -1$)&  $0$   \\
\hline
\multirow{3}{*}{Vector-like leptons} 
&$\mathcal{E}_{L(R)} \equiv [N_1 \ E_1]_{L(R)}$&($\mathbf{1},\mathbf{2}, -1/2$)&  $-(+) 9 Y^\prime / 2$   \\
&$N_{2 L(R)}$&($\mathbf{1},\mathbf{1}, 0$)&  $+(-) 9 Y^\prime / 2$  \\
&$E_{2 L(R)}$&($\mathbf{1},\mathbf{1}, -1$)&  $+(-) 9 Y^\prime / 2$   \\
		\hline \hline
		\multirow{3}{*}{Scalars}
&$\Phi_1$&($\mathbf{1},\mathbf{2}, 1/2$)& $9 Y^\prime$  \\
&$\Phi_2$&($\mathbf{1},\mathbf{2}, 1/2$)& $0$ \\
&$\chi$&($\mathbf{1},\mathbf{1}, 0$)& $9 Y^\prime$  \\
\hline
	\end{tabular}
	\caption{Field content of the type-I UN2HDM and their corresponding transformation properties under the SM and U$(1)^\prime$ symmetries.}
\label{tab:TypeI} 
\end{table*}
The field content of the type-I UN2HDM, first introduced in Ref.~\cite{Aguilar-Saavedra:2022rvy}, is shown in Table~\ref{tab:TypeI}. The SM fermion content is extended with additional VLL doublets $\mathcal{E}_{L,R}$, as well as with charged and neutral VLLs singlets $E_{L,R}$ and $N_{L,R}$. As mentioned before, this extra fields are needed to ensure gauge anomaly cancellation. Furthermore, the VLLs transform non-trivially (with a charge $\omega=e^{i\pi/3}$) under an unbroken $\mathcal{Z}_3$ symmetry, which is introduced to fulfill multiple purposes. Namely, it forbids couplings between the VLLs and SM leptons avoiding any flavor constraints, and also forbids bare Majorana mass terms for the neutral VLLs guaranteeing their Dirac character. Most importantly, the $\mathcal{Z}_3$ stabilizes the lightest VLL which, if neutral, can be a viable DM candidate. The Yukawa Lagrangian involving only the SM fermions is
\begin{equation}
-\mathcal{L}_Y^{\text{I}} = \mathbf{Y}_u \overline{q_L} \tilde{\Phi}_2 u_R + \mathbf{Y}_d \overline{q_L} \Phi_2 d_R + \mathbf{Y}_e \overline{\ell_L} \Phi_2 e_R + {\rm H.c.} \, ,
\label{eq:LYukTypeI}
\end{equation}
with $\mathbf{Y}_{u,d,e}$ being complex Yukawa matrices in generation space. The fermion interactions involving VLLs are
\begin{align}
    - \mathcal{L}_{\rm VLL}^{\text{I}} &= \overline{\mathcal{E}}_L(w_1^N\tilde{\Phi}_2{N_2}_R + w_1^E\Phi_2{E_2}_R) + (w_2^N\overline{N_2}_L\tilde{\Phi}_2^\dagger + w_2^E\overline{E_2}_L\Phi_2^\dagger)\mathcal{E}_R \nonumber \\
    &+ y_2^N\overline{N_2}_L{N_2}_R\chi + y_2^E\overline{E_2}_L{E_2}_R\chi + y_1\overline{\mathcal{E}}_L\mathcal{E}_R\chi^\dagger +{\rm H.c.} \, ,
    \label{eq:YukawasVLLI}
\end{align}
where $w_i^{N,E}$ ($i=1,2$), $y_2^{N,E}$ and $y_1$ are complex Yukawa couplings.

Fermion masses are generated upon spontaneous symmetry breaking~(SSB) by the vacuum expectation values~(VEVs) of the Higgs doublets and singlet, i.e. $\left<\phi^0_{1,2}\right> = v_{1,2}/\sqrt{2}$ and $\left<\chi\right> = u/\sqrt{2}$, respectively (see Sec.~\ref{sec:scalar}). The SM fermions acquire their mass as usual. For the VLLs we obtain the mass Lagrangian
\begin{align}
- \mathcal{L}^{\text{I, mass}}_{\text{VLL}} = \overline{N_L} \mathbf{M}_N N_R + \overline{E_L} \mathbf{M}_E E_R + \text{H.c.} \; ,  \;
\mathbf{M}_{N,E} = \frac{1}{\sqrt{2}} \begin{pmatrix} u y_1& v_2 w_1^{N,E} \\
v_2 w_2^{N,E} & u y_2^{N,E}
\end{pmatrix} \; ,
\end{align}
where $N \equiv (N_1, N_2)$, $E \equiv (E_1, E_2)$ and the $2 \times 2$ mass matrices are written in terms of the Yukawa couplings of Eq.~\eqref{eq:YukawasVLLI}. The above mass matrices are diagonalized via $2 \times 2$ unitary rotations of the neutral and charged VLLs, $N_{L,R} \rightarrow \mathbf{V}_{L,R}^N N_{L,R}$ and $E_{L,R} \rightarrow \mathbf{V}_{L,R}^E E_{L,R}$,
\begin{align}
    \mathbf{V}_L^{N \dagger} \mathbf{M}_N \mathbf{V}_R^{N} = \text{diag}\left(m_{N_1}, m_{N_2} \right) \; , \;
    \mathbf{V}_L^{E \dagger} \mathbf{M}_E \mathbf{V}_R^{E} = \text{diag}\left(m_{E_1}, m_{E_2} \right) \; ,
    \label{eq:VLRTI}
\end{align}
where $m_{N_i}$ and $m_{E_i}$ ($i=1,2$) are, the real and positive, neutral and charged VLL masses, respectively.

\begin{table*}[t!]
\setlength{\tabcolsep}{-1pt}
\centering
\setlength{\tabcolsep}{20pt}
\begin{tabular}{|c|c|c|c|}
		\hline
 &Fields&$G_{\rm SM}$&    U($1$)$^\prime$  \\ 
		\hline \hline
		\multirow{5}{*}{SM fermions}
&$q_L$&($\mathbf{3},\mathbf{2}, 1/6$)& $0$ \\
&$u_R$&($\mathbf{3},\mathbf{1}, 2/3$)&  $Y^\prime$   \\
&$d_R$&($\mathbf{3},\mathbf{1}, -1/3$)&  $-Y^\prime$  \\
&$\ell_L$&($\mathbf{1},\mathbf{2}, -1/2$)&  $0$   \\
&$e_R$&($\mathbf{1},\mathbf{1}, -1$)&  $0$   \\
\hline
\multirow{2}{*}{Vector-like leptons} &$N_{1,2,3 L (R)}$&($\mathbf{1},\mathbf{1}, 0$)&  0 $(Y^\prime)$   \\
&$E_{1,2,3 L (R)}$&($\mathbf{1},\mathbf{1}, -1$)&  $0$ $(-Y^\prime)$  \\
		\hline \hline
		\multirow{3}{*}{Scalars}
&$\Phi_1$&($\mathbf{1},\mathbf{2}, 1/2$)& $0$  \\
&$\Phi_2$&($\mathbf{1},\mathbf{2}, 1/2$)& $Y^\prime$  \\
&$\chi$&($\mathbf{1},\mathbf{1}, 0$)& $-Y^\prime$  \\
\hline
	\end{tabular}
	\caption{Field content of the type-III/lepton specific UN2HDM and their corresponding transformation properties under the SM  and U$(1)^\prime$ symmetries.}
	\label{tab:TypeIII} 
\end{table*}
The type-III (or lepton-specific) UN2HDM is realized via the field content and charge assignments shown in Table~\ref{tab:TypeIII}. Note that three generations of $N_{L,R}$ and $E_{L,R}$ VLL singlets are now needed for anomaly cancellation. As in the type-I scenario, the VLLs transform non-trivially under a $\mathcal{Z}_3$ symmetry with a charge $\omega=e^{i\pi/3}$. The Yukawa Lagrangian for the SM fermion fields is
\begin{align}
-\mathcal{L}_Y^{\text{III}} = \mathbf{Y}_u \overline{q_L} \tilde{\Phi}_2 u_R + \mathbf{Y}_d \overline{q_L} \Phi_2 d_R + \mathbf{Y}_e \overline{\ell_L} \Phi_1 e_R + {\rm H.c.}\,,
\end{align}
where $\mathbf{Y}_{u,d,e}$ are $3 \times 3$ complex Yukawa matrices, while for the VLLs
\begin{align}
    -\mathcal{L}_{\rm VLL}^{\text{III}} &= \mathbf{Y}_N \overline{N_L} {N}_R \chi + \mathbf{Y}_E \overline{E_L} {E}_R \chi^\ast + {\rm H.c.}\,,
    \label{eq:YukawasVLLIII}
\end{align}
with $\mathbf{Y}_{N,E}$ being $3 \times 3$ complex Yukawa matrices, $N \equiv (N_1, N_2, N_3)$ and $E \equiv (E_1, E_2, E_3)$. After SSB, the VLLs mass Lagrangian is
\begin{align}
- \mathcal{L}^{\text{III, mass}}_{\text{VLL}} = \overline{N_L} \mathbf{M}_N N_R + \overline{E_L} \mathbf{M}_E E_R + \text{H.c.} \; , \;
\mathbf{M}_{N,E} = \frac{u \mathbf{Y}_{N,E}}{\sqrt{2}} \; ,
\end{align}
where the $3 \times 3$ mass matrices above are diagonalized via $3 \times 3$ unitary rotations, $N_{L,R} \rightarrow \mathbf{V}_{L,R}^N N_{L,R}$ and $E_{L,R} \rightarrow \mathbf{V}_{L,R}^E E_{L,R}$ as
\begin{align}
    \mathbf{V}_L^{N \dagger} \mathbf{M}_N \mathbf{V}_R^{N} = \text{diag}\left(m_{N_1}, m_{N_2}, m_{N_3} \right) \; , \;
    \mathbf{V}_L^{E \dagger} \mathbf{M}_E \mathbf{V}_R^{E} = \text{diag}\left(m_{E_1}, m_{E_2}, m_{E_3} \right) \; .
\end{align}

\subsection{Scalar sector}
\label{sec:scalar}

The scalar potential of the UN2HDM is given by
\begin{align}
V &= m_{11}^2\Phi_1^\dagger\Phi_1 + m_{22}^2\Phi_2^\dagger\Phi_2 + \frac{m_0^2}{2}\chi^\dagger\chi + \frac{\lambda_1}{2}(\Phi_1^\dagger\Phi_1)^2 + \frac{\lambda_2}{2}(\Phi_2^\dagger\Phi_2)^2 
+ \lambda_3(\Phi_1^\dagger\Phi_1)(\Phi_2^\dagger\Phi_2) \notag \\
& + \lambda_4(\Phi_1^\dagger\Phi_2)(\Phi_2^\dagger\Phi_1) + \frac{\lambda_5}{2}(\chi^\dagger\chi)^2 + \frac{\lambda_6}{2}(\Phi_1^\dagger\Phi_1)(\chi^\dagger\chi) + \frac{\lambda_7}{2}(\Phi_2^\dagger\Phi_2)(\chi^\dagger\chi) + (\,\mu\chi\Phi_1^\dagger\Phi_2+ {\rm H.c.})\,,
\label{eq:V}
\end{align}
where $\mu$ can be made real. We define the scalar doublets $\Phi_{1,2}$ and singlet $\chi$ as 
\begin{align}
    \Phi_{1,2} = \begin{pmatrix} \phi_{1,2}^+ \\ \phi^0_{1,2} \end{pmatrix} = \frac{1}{\sqrt{2}}\begin{pmatrix} \sqrt{2}\phi_{1,2}^+ \\ v_{1,2} + \rho_{1,2} + i\eta_{1,2} \end{pmatrix} \; , \;
    \chi = \frac{1}{\sqrt{2}}(u + \rho_3 + i\eta_3)\,,
\end{align}
and, without loss of generality, assume that the VEVs are real
\begin{equation}
    \begin{gathered}
    \langle\Phi_{1,2}\rangle = \frac{1}{\sqrt{2}}\begin{pmatrix} 0 \\ v_{1,2} \end{pmatrix}\,, \quad \langle\chi\rangle = \frac{1}{\sqrt{2}}u \; , \; v = \sqrt{v_1^2 + v_2^2} \simeq 246 \ \text{GeV} \quad,\quad \tan\beta = v_2/v_1\,.
    \end{gathered}
    \label{eq:VEV}
\end{equation}
% %
% where,
% %
% \begin{equation}
% v = \sqrt{v_1^2 + v_2^2} \simeq 246 \ \text{GeV} \quad,\quad \tan\beta = v_2/v_1\,.
% \label{eq:tanbeta}
% \end{equation}
%

For $v_1,v_2,u\neq 0$, one obtains three minimisation conditions which allow to write $m_{11}^2$, $m_{22}^2$ and $m_0^2$ in terms of the VEVs and the remaining parameters of $V$. The $6\times 6$ neutral scalar mass matrix can be written as
\begin{equation}
\mathcal{M}^n = \begin{pmatrix} M^\rho & 0 \\ 0 & M^\eta \end{pmatrix}\,,
\end{equation}
where $M^\rho$ and $M^\eta$ are $3\times 3$ real symmetric matrices defined in the $(\rho_1,\rho_2,\rho_3)$ and $(\eta_1,\eta_2,\eta_3)$ bases, respectively. The elements of $M^\rho$ read
\begin{align}
& M^\rho_{11} = v^2\lambda_1\cbeta^2 - \frac{1}{\sqrt{2}}u\mu\tan\beta\;, \; M^\rho_{12} = v^2(\lambda_3+\lambda_4)\sbeta\cbeta + \frac{1}{\sqrt{2}}u\mu\;, \; M^\rho_{13} = \frac{v}{2}u\lambda_6\cbeta + \frac{1}{\sqrt{2}}v\mu\sbeta\;, \notag \\
& M^\rho_{22} = v^2\lambda_2\sbeta^2 - \frac{u\mu}{\sqrt{2}\tan\beta}\;, \; M^\rho_{23} = \frac{v}{2}u\lambda_7\sbeta + \frac{1}{\sqrt{2}}v\mu\cbeta\;, \; M^\rho_{33} = \lambda_5u^2 - \frac{1}{\sqrt{2}}\frac{v^2}{u}\mu\cbeta\sbeta\;, 
\label{eq:Mrho}
\end{align}
where we use the shorthand notation $s_\beta = \sin\beta$, $c_\beta = \cos\beta$. For $M_\eta$ we have
\begin{align}
M^\eta_{11} &= -\dfrac{1}{\sqrt{2}}u\mu\dfrac{\sbeta}{\cbeta} \; , \;
 M^\eta_{12} = \dfrac{1}{\sqrt{2}}u\mu \; , \; M^\eta_{13} = \dfrac{1}{\sqrt{2}}v\mu\sbeta \; , \nonumber \\
 M^\eta_{22} & = -\frac{1}{\sqrt{2}}u\mu\dfrac{\cbeta}{\sbeta} \; , \; M^\eta_{23}  = -\dfrac{1}{\sqrt{2}}v\mu\cbeta \; , \;
 M^\eta_{33} = -\frac{1}{\sqrt{2}}\dfrac{v^2}{u}\mu\cbeta\sbeta \,. 
\label{eq:Meta}
\end{align}
The $3 \times 3$ orthogonal matrix $O$ that rotates the $\rho_i$ fields into the mass basis, can be parametrized by three mixing angles, $\alpha_1$, $\alpha_2$, $\alpha_3$,
\begin{equation}    
O = \begin{pmatrix} c_1c_2 & s_1c_2 & s_2 \\ -c_1s_2s_3 - s_1c_3 & c_1c_3 - s_1s_2s_3 & c_2s_3 \\ -c_1s_2c_3 + s_1s_3 & -c_1s_3 - s_1s_2c_3 & c_2c_3 
\end{pmatrix}^T\,,
\label{eq:O}
\end{equation}
where $c_{1,2,3}=\cos\alpha_{1,2,3}$, $s_{1,2,3}=\sin\alpha_{1,2,3}$ and $-\pi/2 \leq \alpha_{1,2,3} \leq \pi/2$. In what follows, $h$ corresponds to the SM Higgs boson with mass $m_h=125.09$~GeV~\cite{ParticleDataGroup:2022pth} and $H_{1,2}$ are the new CP-even scalars, with masses $m_{H_1}<m_{H_2}$. We can write $M^\rho$ in terms of the three scalar masses and mixing angles introduced in Eq.~(\ref{eq:O}). Namely,
\begin{equation}
M^\rho = O\begin{pmatrix} m_h^2 & 0 & 0 \\ 0 & m_{H_1}^2 & 0 \\ 0 & 0 & m_{H_2}^2 \end{pmatrix}O^T\,.
\label{eq:MH}
\end{equation}
The matrix $M^\eta$ is diagonalized as $R^T M^\eta R = (M^\eta)_\text{diag}$, using an orthogonal rotation
\begin{equation}
R = \left( \begin{array}{ccc}
-\sbeta & \cbeta & 0 \\ \cbeta & \sbeta & 0 \\ 0 & 0 & 1
\end{array}
\right) 
 \left( \begin{array}{ccc}
-\salpha & 0 & \calpha  \\ 0 & 1 & 0 \\ \calpha & 0 & \salpha
\end{array}
\right) \; , \;  \tan \alpha = -\frac{u}{v\cbeta\sbeta} \,.
\label{eq:Rrot}
\end{equation}
The only non-zero eigenvalue of $M^\eta$ is
\begin{equation}
m_{A^0}^2 = - \frac{\mu}{\sqrt{2}} \left[\frac{u}{\sbeta \cbeta} + \frac{\sbeta \cbeta v^2}{u} \right]\,,
\label{eq:mA0}
\end{equation}
which corresponds to the squared mass of a pseudoscalar~$A^0$. Finally, the charged-scalar mass matrix in the basis $(\phi_1^\pm,\phi_2^\pm)$ is
\begin{equation}
    \mathcal{M}^c = \left(v^2\frac{\lambda_4}{2} + \frac{\sqrt{2}u\mu}{2 s_\beta c_\beta}\right)
    \begin{pmatrix} -s_\beta^2 & s_\beta c_\beta \\
    s_\beta c_\beta & -c_\beta^2 \end{pmatrix}\,,
\end{equation}
and is diagonalized as $U^T \mathcal{M}^c U = (\mathcal{M}^c)_\text{diag}$, with
\begin{equation}
U = \left( \begin{array}{cc}
-\sbeta & \cbeta \\ \cbeta & \sbeta
\end{array}
\right)  \,,
\label{eq:U}
\end{equation}
as in the usual 2HDM~\cite{Branco:2011iw}. The non-zero eigenvalue of $\mathcal{M}^c$ is
\begin{equation}
    m_{H^\pm}^2 = -v^2 \frac{\lambda_4}{2} - \frac{\sqrt{2}u\mu}{2 \cbeta\sbeta}\,,
    \label{eq:Mch}
\end{equation}
corresponding to the squared mass of the new charged scalar~$H^\pm$. 

An important feature of $V$ is that the number of parameters is equal to the number of physical quantities (masses and mixing angles) needed to define the scalar sector. Thus, all eleven parameters shown in Eq.~(\ref{eq:V}) can be written in terms of the three VEVs, as well as of the five physical scalar masses and three mixing angles, which we introduced above. In Appendix~\ref{sec:parameters} we show how the scalar potential parameters can be written as functions of the physical parameters we have just presented.

%%%%%%%%%%%%%%%%%%%%%%%%%%%%%%%%%%%%%%%%%%%%%%%%%%%%%%%%%%%%%%%%%%%%%%%%%%%%%
\section{Numerical procedure and results}
\label{sec:numerical}
%%%%%%%%%%%%%%%%%%%%%%%%%%%%%%%%%%%%%%%%%%%%%%%%%%%%%%%%%%%%%%%%%%%%%%%%%%%%%

%
\begin{table}[!t]
\centering
\begin{tabular}{cc}  
\hline
 Parameters & Scan range \\
\hline
$m_{H_1}$ & $[94 , 97]$ GeV \\
$m_h$ & $125.09$ GeV \\
$m_{H_2}$ & $[125.1 , 1000]$ GeV \\
$m_{A^0} , m_{H^{\pm}}$ & $[20 , 1000]$ GeV \\
$\tan \beta$ & $[0, 20]$ \\
$c(hVV)$ & $[0.9,1.0]$ \\
$c(h t \overline{t})$ &  $[0.8,1.2]$ \\
sign$(O_{31})$ & $\{-1,1\}$\\
$O_{32}$ &  $[-1,1]$ \\
Yukawa type & $\{{\rm I,III}\}$ \\
$m_{Z^\prime}$ & $2.2$ TeV \\
$g_{Z^\prime} Y^\prime$ &  $0.1$ \\
$m_E$ &  $[300,5000]$ GeV \\
\hline
\end{tabular}

\caption{Input parameters and corresponding ranges used in our numerical scan.}
\label{tab:Scan}
\end{table}
We now explore the UN2HDM parameter space to investigate whether the decays of a 95~GeV neutral scalar could explain the $\gamma\gamma$, $\tau\tau$ and/or $b\bar{b}$ excesses, and be consistent with other experimental data. To this end, we perform numerical scans varying the model parameters as shown in Table~\ref{tab:Scan}. We assume $H_2$ to be always the heaviest CP-even neutral scalar, while the mass of $H_1$ is allowed to vary in a small interval around 95~GeV. These masses, together with sign($O_{31}$) and $O_{32}$, the effective couplings of the SM-like Higgs boson $h$ to SM gauge bosons and top quarks, respectively denoted as $c(hVV)$ ($V=W,Z$) and $c(ht\overline{t})$, parametrize the mixing matrix of CP-even scalars given in Eq.~(\ref{eq:O}). This choice of parameters increases the efficiency of the scan, since it allows us to compute the three mixing angles of $O$ while simultaneously imposing SM-like couplings for $h$. The mass of $Z^\prime$ ($m_{Z^\prime}$) and the product $g_{Z^\prime} Y^\prime$ are required to extract the VEV of the scalar singlet $u$, as well as the $Z-Z'$ mixing angle $\theta_Z$. We use values of $m_{Z^\prime}$ and $g_{Z^\prime} Y^\prime$ similar to those used in~\cite{Aguilar-Saavedra:2022rvy}, with the former being motivated by the 2~TeV \emph{bumps} reported by ATLAS and CMS~\cite{ATLAS:2015xom, ATLAS:2017zuf, CMS:2017fgc, arXiv:2210.00043}. Considering the inputs of Table~\ref{tab:Scan}, we get $u\sim 2.4$~TeV ($u\sim 20$~TeV) for the type-I (III) UN2HDM. In all cases, the $Z-Z'$ mixing angle is small enough ($\theta_Z<10^{-3}$) to fulfill the constraints coming from electroweak precision data~\cite{Erler:2009jh}.

The generated points are passed through the \texttt{ScannerS-2}~\cite{Muhlleitner:2020wwk} code, to which the type of Yukawa sector (I or III) is also provided. At this point, the following constraints are applied:

\begin{itemize}
\item \textbf{Perturbative unitarity:} \texttt{ScannerS-2} makes use of the analytical results from Ref.~\cite{Muhlleitner:2016mzt} to guarantee that the eigenvalues of the $2-2$ (pseudo)scalar-(pseudo)scalar scattering matrix are below $8 \pi$ and, thus, ensure tree-level perturbative unitarity.

\item \textbf{Vacuum stability:} boundness from below of the scalar potential is checked by  \texttt{ScannerS-2}, which implements the conditions obtained in Ref.~\cite{Muhlleitner:2016mzt} and simultaneously provides a link to the library \texttt{EVADE}~\cite{Hollik:2018wrr, Ferreira:2019iqb} to verify if a certain point corresponds to the global minimum of the potential. 

\item \textbf{Electroweak precision observables~(EWPO):} Singlet VLLs do not contribute to the oblique parameters. This is also the case for the doublet VLL (type-I UN2HDM), in the considered limit of no mixing and degenerate masses. Following Refs.~\cite{Grimus:2007if, Grimus:2008nb}, where analytical expressions for $S$, $T$ and $U$ in a general SM extension with an arbitrary number of scalar doublets and singlets were derived, \texttt{ScannerS-2} is able to compute those parameters in our model. These are compared with the global fit results from Ref.~\cite{Haller:2018nnx}, guaranteeing compatibility at the $95 \%$ confidence level (CL). The values used for the oblique parameters are:
\begin{equation}
S=0.04 \pm 0.11 \; , \; T=0.09\pm0.14 \; , \; U=-0.02\pm0.11 \; ,
\end{equation}
with correlation coefficients of $+0.92$ ($-0.68$) [$-0.87$] between $S$ and $T$ ($S$ and $U$) [$T$ and $U$]. %{\red put the ranges somewhere here}

\item \textbf{Flavor observables:} Besides EWPO, the global-fit results of Ref.~\cite{Haller:2018nnx} also include constraints coming from $B$-physics observables, setting limits on the $(m_{H^\pm}, \tan\beta)$ plane. These are taken into account in our scan. Depending on the Yukawa sector type, some flavor observables will be more constraining than others. Namely, for the type-I and III Yukawa, the most relevant process is $B_d \rightarrow \mu^+ \mu^-$.
\end{itemize}

For all points passing the aforementioned constraints, the branching ratios~(BRs) of the UN2HDM scalars are computed by \texttt{ScannerS-2}, where an interface to a modified version of the code \texttt{N2HDECAY}~\cite{Engeln:2018mbg} is used. Namely, due to differences between the scalar potential of the UN2HDM [see Eq.~\eqref{eq:V}] and the one of Ref.~\cite{Engeln:2018mbg}, we apply the necessary changes to the triple scalar couplings. Furthermore, given that $\theta_Z\ll 1$, we neglect the terms arising from $Z-Z'$ mixing in the couplings between scalars and SM gauge bosons. Since \texttt{N2HDECAY} does not include the contribution of the charged VLLs to the Higgs-to-diphoton BR, we compute it analytically at one-loop level (see Appendix~\ref{sec:diphoton}). We make the VLL mass $m_{E}$ vary within the ranges shown in Table~\ref{tab:Scan}. Lastly, in order to extract the signal strengths we make use of \texttt{HiggsPredictions} to obtain the relevant cross-sections. This code is part of \texttt{HiggsTools}~\cite{Bahl:2022igd}, which also comprises the latest versions of \texttt{HiggsBounds}~\cite{Bechtle:2008jh, Bechtle:2011sb, Bechtle:2013wla, Bechtle:2015pma, Bechtle:2020pkv} and \texttt{HiggsSignals}~\cite{Bechtle:2013xfa, Bechtle:2020uwn}. For the viable \texttt{ScannerS-2} outputs, we incorporate the following experimental constraints:

\begin{itemize}
\item \textbf{125 GeV Higgs boson measurements:} The set of generated points must be compatible with the experimentally measured properties of the SM-like Higgs boson with mass 125~GeV. This is done using \texttt{HiggsSignals}, which provides $\chi^2$ values reflecting the agreement of the model predictions with the measurements performed at the LHC. We define the following quantity, 
\begin{equation}
\Delta\chi^2_{125} = \chi^2 - \chi^2_{\rm SM}\,,
\label{eq:chi125}
\end{equation}
where $\chi^2_{\rm SM}=152.5$, is the $\chi^2$ value of the SM provided by \texttt{HiggsSignals}, for 159 degrees of freedom. This represents an agreement within a 37\% CL. In our work we use two criteria which we dub as \emph {tight} and \emph{loose}. The tight constraint requires $\Delta\chi^2_{125} \leq 6.18$, with the precise value chosen for better comparison with Ref.~\cite{Biekotter:2023oen}. This represents agreement 
with experimental data at the 50.7\% CL (still better than $1\sigma$). Instead, the loose constraint requires agreement at the 95\% CL, that is, $\Delta\chi^2_{125} \leq 37$.

\item \textbf{New scalar searches:} The parameter space of our model is further constrained by searches for potential new scalars performed at LEP and LHC. This analysis is realized thanks to \texttt{HiggsBounds}. Note that we also take into account a recent CMS search~\cite{CMS-PAS-EXO-21-018} that is currently not included in \texttt{HiggsBounds}.
\end{itemize}

Finally, the strength parameters $\mu_{\gamma \gamma}$, $\mu_{\tau \tau}$ and $\mu_{b \bar b}$ are computed for the scalar $H_1$ with mass around $95$~GeV. They are defined as
\begin{align}
    \mu_{\gamma \gamma} &= \frac{\sigma(g g \rightarrow H_1 \rightarrow \gamma \gamma)}{\sigma(g g \rightarrow H_1 \rightarrow \gamma \gamma)_{\text{SM}}}
    %=  \frac{\sigma(g g \rightarrow H_1)}{\sigma(g g \rightarrow H_1)_{\text{SM}}} \times \frac{\text{BR}(H_1 \rightarrow \gamma \gamma)}{\text{BR}(H_1 \rightarrow \gamma \gamma)_{\text{SM}}}
    \; , \nonumber \\
    \mu_{\tau \tau} &= \frac{\sigma(g g \rightarrow H_1 \rightarrow \tau \tau)}{\sigma(g g \rightarrow H_1 \rightarrow \tau \tau)_{\text{SM}}} 
    %=  \frac{\sigma(g g \rightarrow H_1)}{\sigma(g g \rightarrow H_1)_{\text{SM}}} \times \frac{\text{BR}(H_1 \rightarrow \tau \tau)}{\text{BR}(H_1 \rightarrow \tau \tau)_{\text{SM}}}
    \; , \nonumber \\
    \mu_{b \bar b} &=  \frac{\sigma(e^+ e^- \rightarrow Z H_1 \rightarrow b \overline{b})}{\sigma(e^+ e^- \rightarrow Z H_1 \rightarrow b \overline{b})_{\text{SM}}}
    %=  \frac{\sigma(e^+ e^- \rightarrow Z H_1)}{\sigma(e^+ e^- \rightarrow Z H_1)_{\text{SM}}} \times \frac{\text{BR}(H_1 \rightarrow b \overline{b})}{\text{BR}(H_1 \rightarrow b \overline{b})_{\text{SM}}}
    \; ,
    \label{eq:mudef}
\end{align}
where the subscript SM refers to quantities computed within the SM for a Higgs-like scalar of mass $m_{H_1}$. 

\begin{figure}[t!]
\begin{center}
\begin{tabular}{ccc}
\includegraphics[height=6.8cm,clip=]{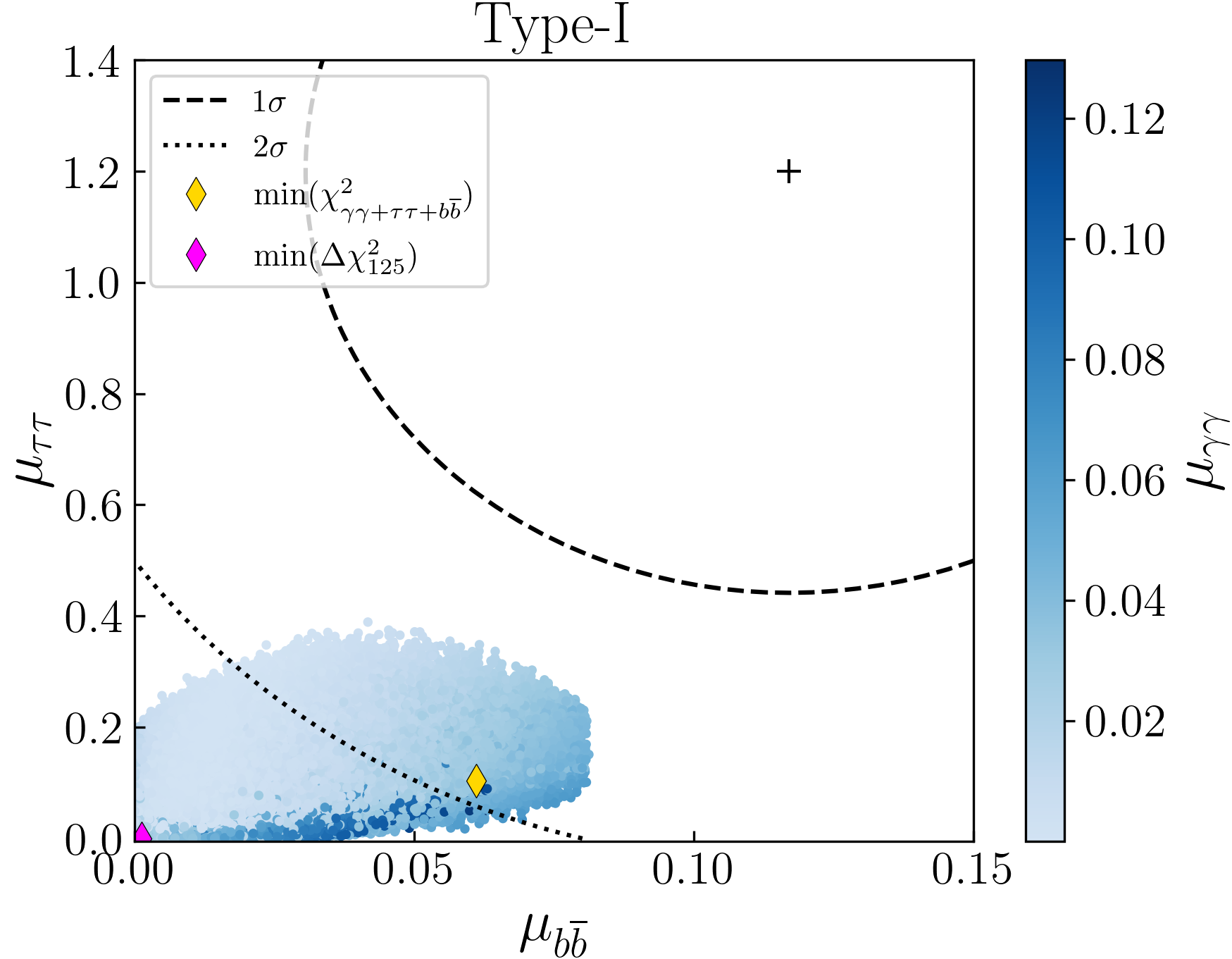} & \quad &
\includegraphics[height=6.8cm,clip=]{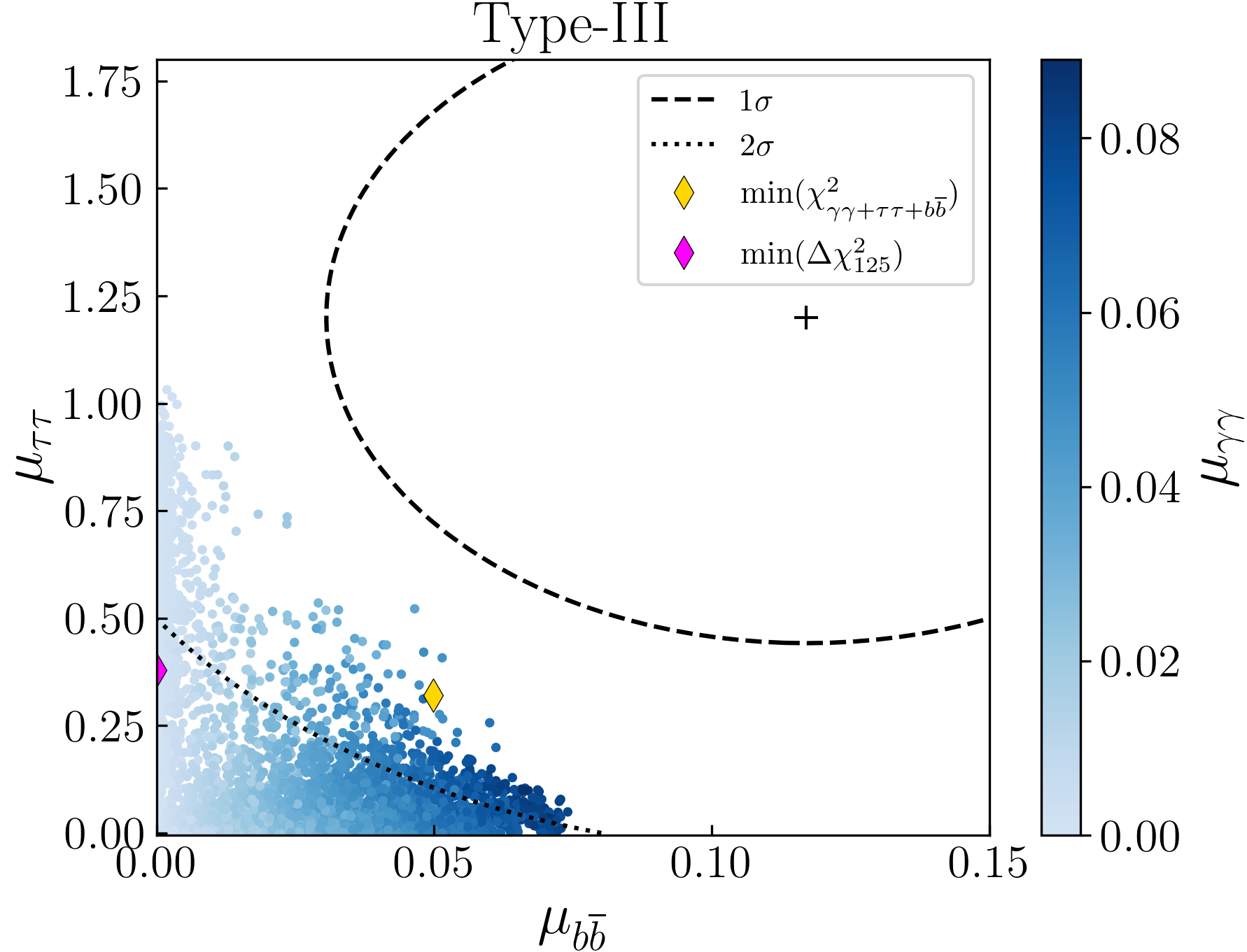}
\end{tabular}
\end{center}
\caption{Allowed values for the strength parameters for the type-I (left) and type-III (right) UN2HDM. The crosses represent the best-fit values for $(\mu_{b \bar b},\mu_{\tau \tau})$ and the dashed (dotted) contours represent $1 \sigma$ ($2 \sigma$) CL regions. The color code indicates the value of $\mu_{\gamma \gamma}$. The yellow (purple) diamond are at the minimum value for $\chi^2_{\gamma \gamma + \tau \tau + b \bar b}$ ($\Delta \chi^2_{\text{125}}$) (see the text).}
\label{fig:UN2HDMTImus}
\end{figure}

The allowed regions for $\mu_{b \bar b}$, $\mu_{\tau \tau}$ and $\mu_{\gamma \gamma}$ imposing the tight constraint on $\Delta \chi_{125}^2$ are presented in Fig.~\ref{fig:UN2HDMTImus} for the type-I and type-III UN2HDM, where the color grading indicates the value of $\mu_{\gamma \gamma}$, which in all cases is well below the experimental best-fit value $\mu_{\gamma \gamma}^\text{exp} = 0.27$. 
The purple diamond represents the point with smallest $\chi^2_{\text{125}}$, while the yellow one indicates the minimum value for $\chi^2_{\gamma \gamma + \tau \tau + b \bar b} \equiv \chi^2_{\gamma \gamma} + \chi^2_{\tau \tau} + \chi^2_{b \bar b}$, where
\begin{equation}
    \chi^2_{{\gamma \gamma, \tau \tau, b \bar b}} = \left(\frac{\mu_{\gamma \gamma, \tau \tau, b \bar b} - \mu_{\gamma \gamma, \tau \tau, b \bar b}^{\text{exp}}}{\Delta \mu_{\gamma \gamma, \tau \tau, b \bar b}^{\text{exp}}}\right)^2 \,,
    \label{eq:chimu}
\end{equation}
being $\Delta \mu_{\gamma \gamma, \tau \tau, b \bar b}^{\text{exp}}$ the experimental uncertainty of $\mu_{\gamma \gamma, \tau \tau, b \bar b}^{\text{exp}}$ given in Eqs.~\eqref{eq:mugg} and \eqref{eq:muttbb}.

\begin{figure}[p]
\begin{center}
\begin{tabular}{ccc}
\includegraphics[height=6.5cm,clip]{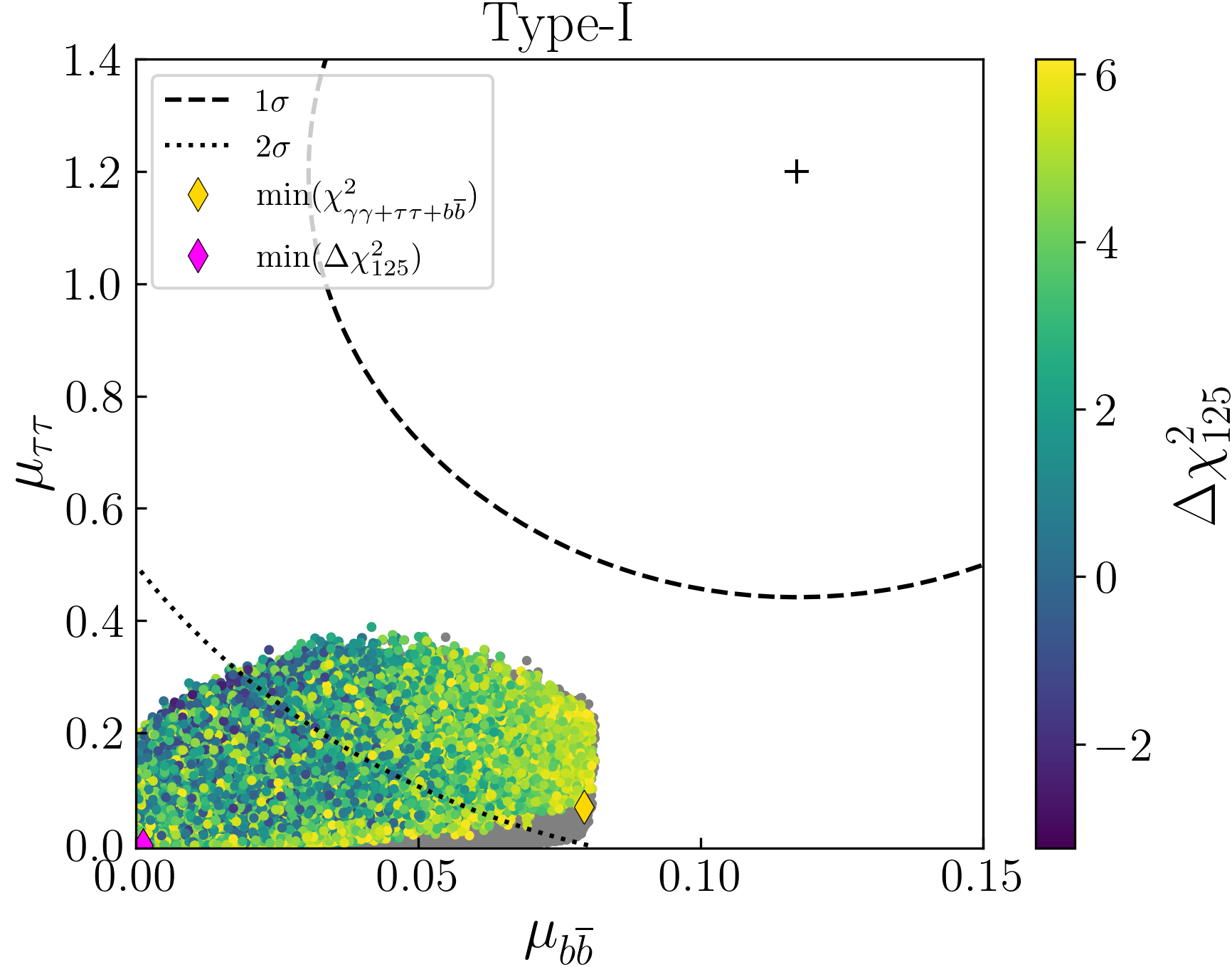} & &
\includegraphics[height=6.5cm,clip]{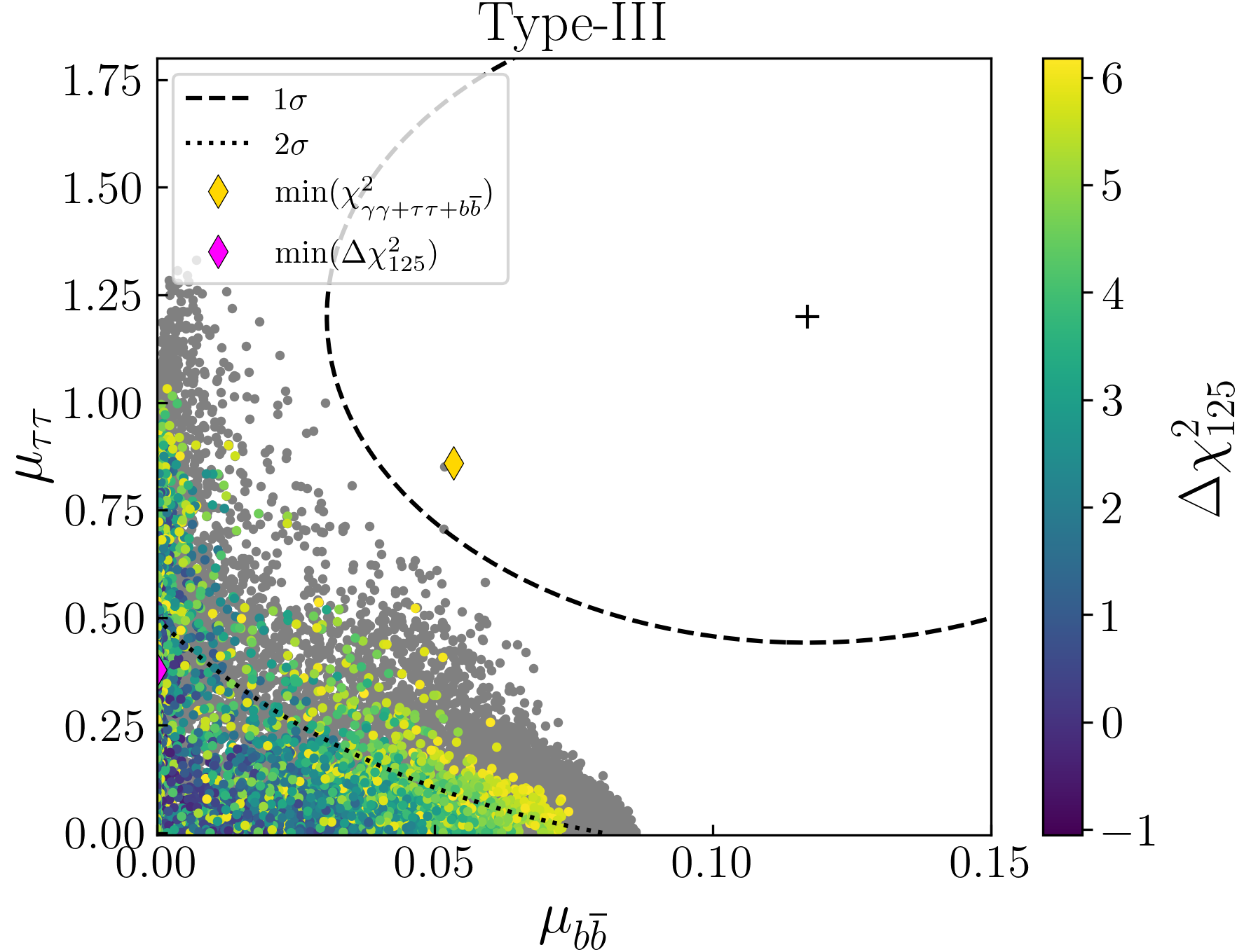}\\[2mm]
\includegraphics[height=6.5cm,clip]{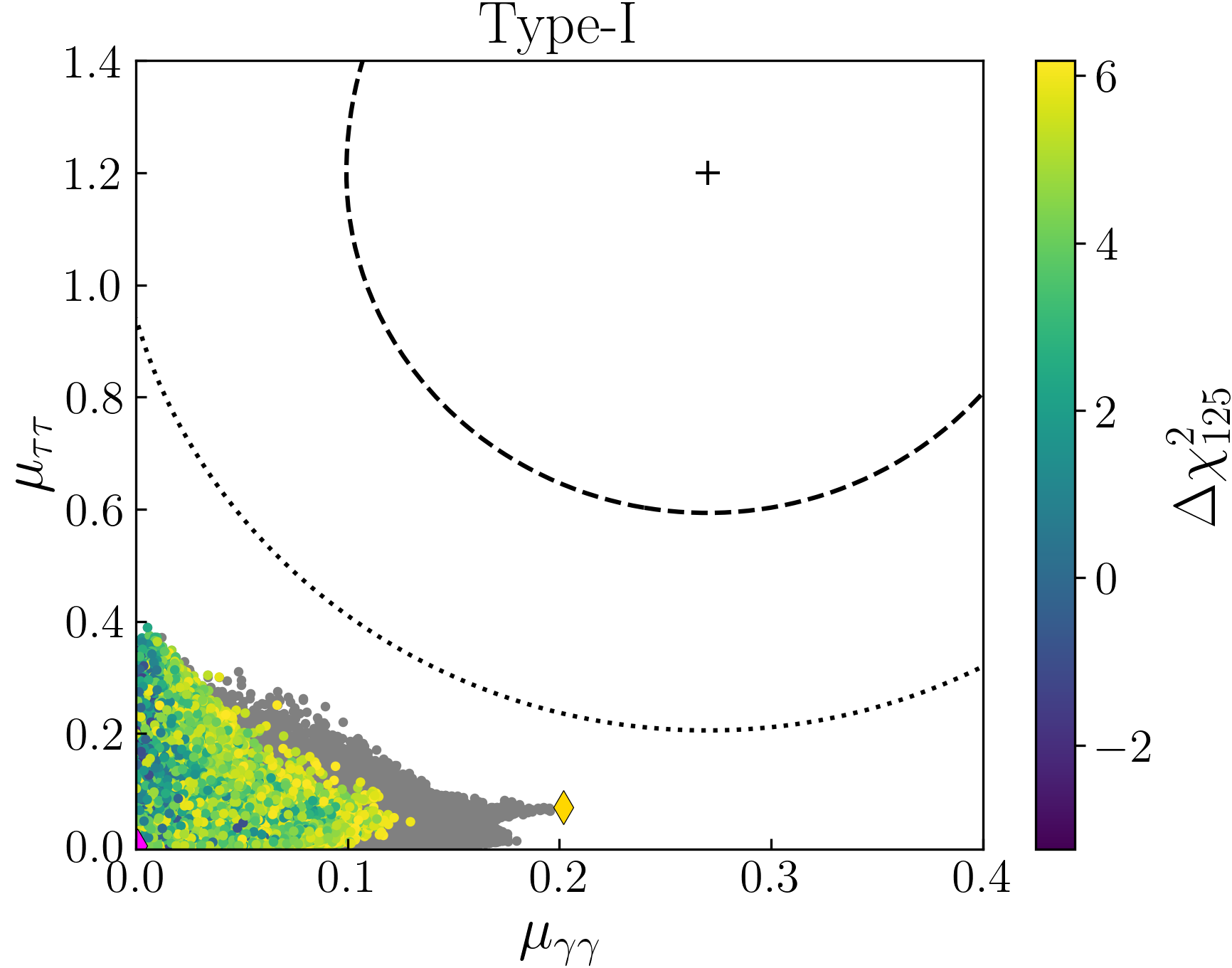} & & 
\includegraphics[height=6.5cm,clip]{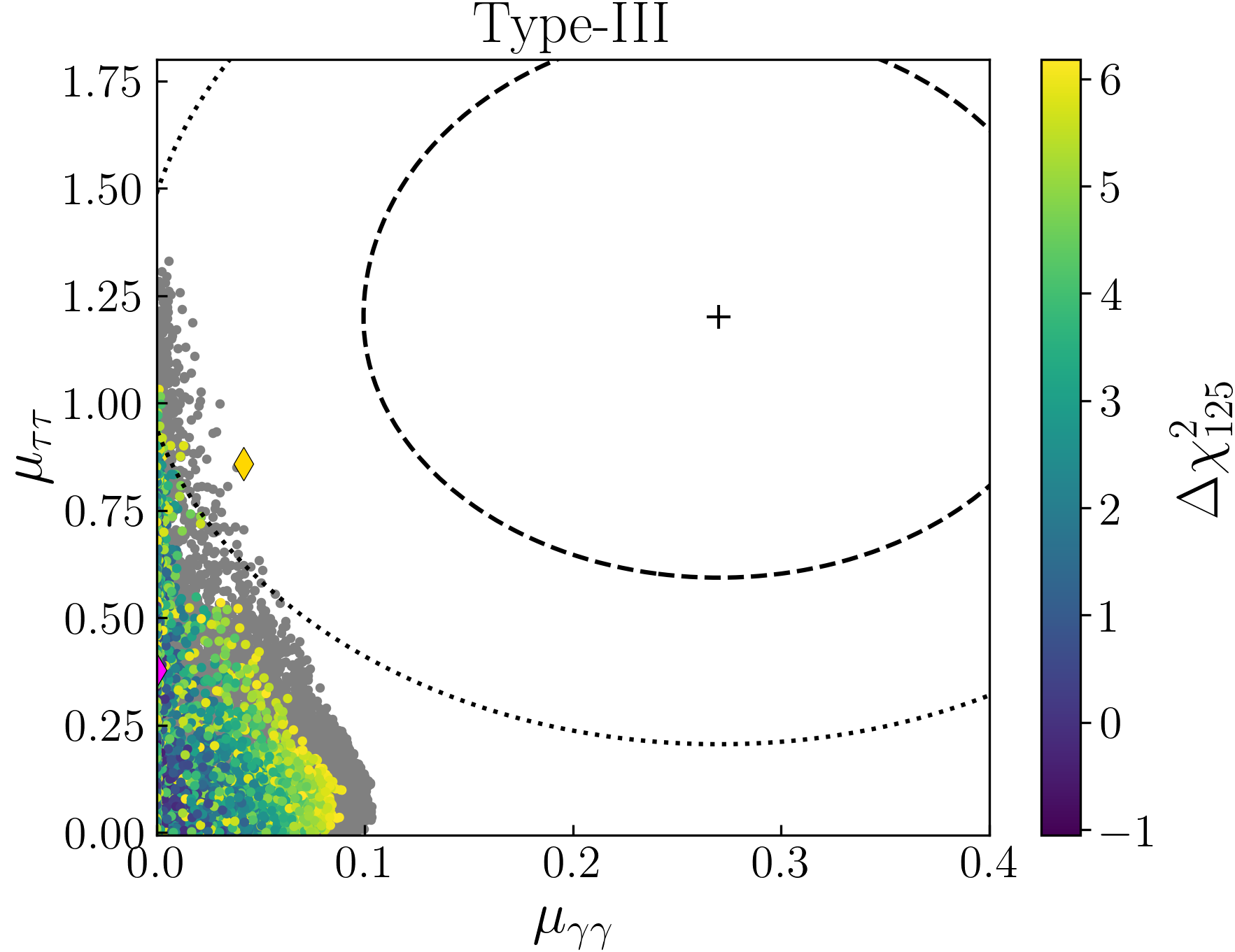} \\[2mm]
\includegraphics[height=6.5cm,clip]{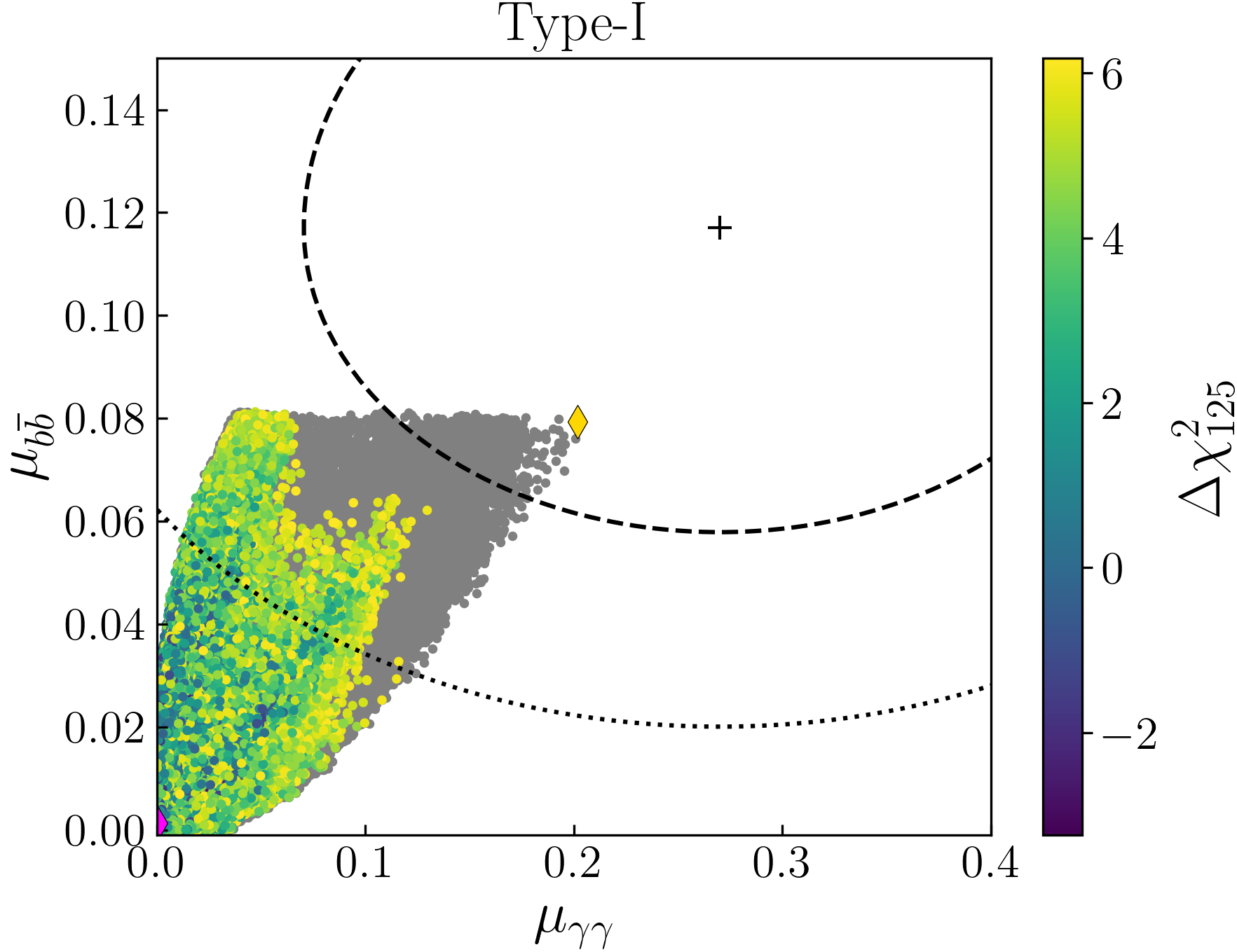} & &
\includegraphics[height=6.5cm,clip]{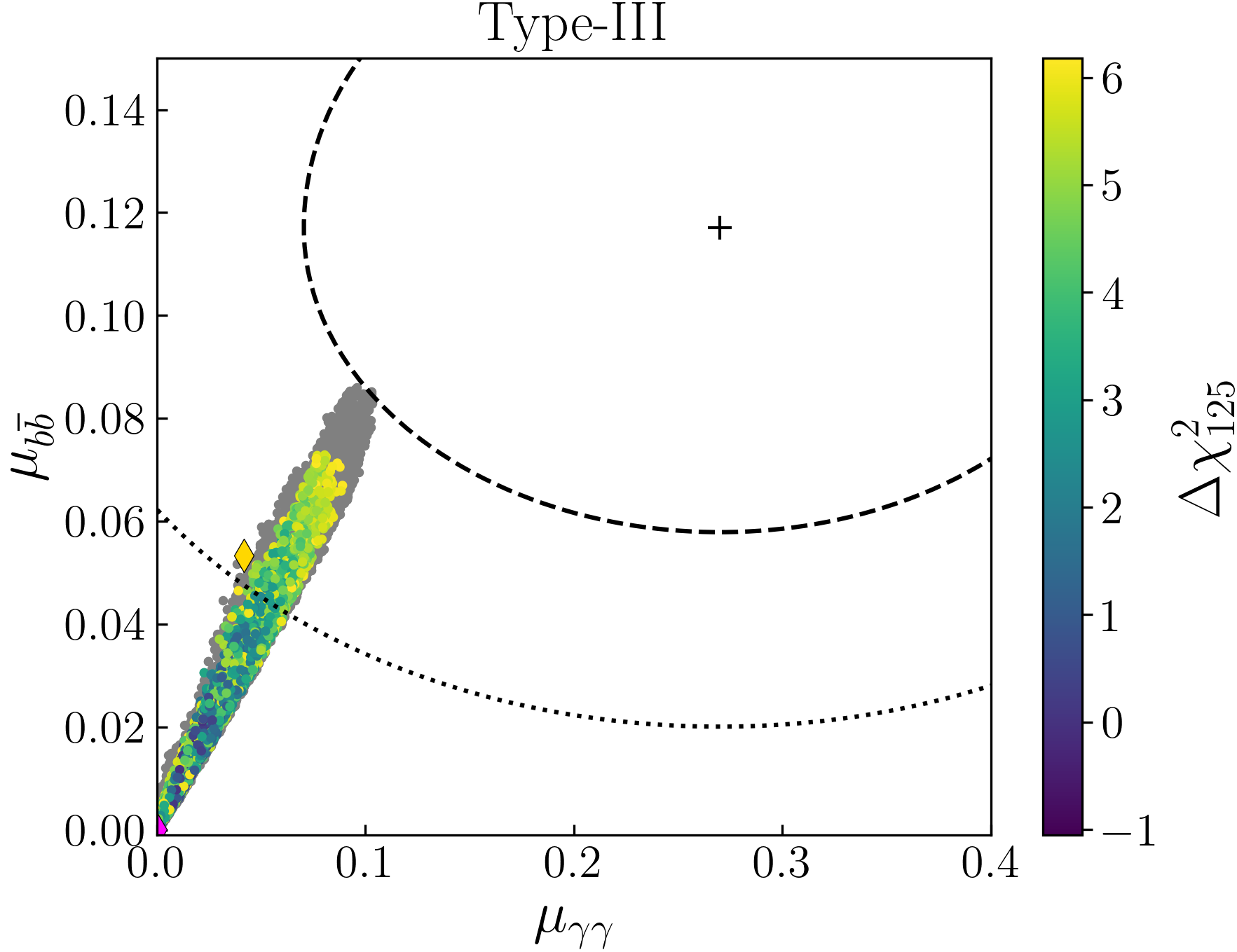} \\
\end{tabular}
\end{center}
\caption{Allowed values for the strength parameters for the type-I (left) and type-III (right) UN2HDM, in the $(\mu_{b \bar b},\mu_{\tau \tau})$ (top), $(\mu_{\gamma \gamma},\mu_{\tau \tau})$ (middle) and $(\mu_{\gamma \gamma},\mu_{b \bar b})$ (bottom) planes. The crosses represent the best-fit values for the strength parameters and the dashed (dotted) contours represent $1 \sigma$ ($2 \sigma$) CL regions. The color code indicates the value of $\Delta \chi_{125}^2$ and the yellow (purple) diamond indicates the minimum value for $\chi^2_{\gamma \gamma + \tau \tau + b \bar b}$ ($\Delta\chi^2_{\text{125}}$) (see the text).}
\label{fig:UN2HDMTIchi2}
\end{figure}

Relaxing the constraint on $\Delta \chi_{125}^2$, the allowed regions for $\mu_{b \bar b}$, $\mu_{\tau \tau}$ and $\mu_{\gamma \gamma}$ are presented in Fig.~\ref{fig:UN2HDMTIchi2}. Here, the color grading represents $\Delta \chi_{125}^2$ when $\Delta\chi^2_{125} \leq 6.18$, and the gray points are those with $6.18 < \Delta\chi^2_{125} \leq 37$. The loose constraint does not significantly enlarge the regions in the $(\mu_{b \bar b},\mu_{\tau \tau})$ plane (top panels) --- note that for type-III the allowed region has a similar size but is more populated. On the other hand, the loose constraint allows for slightly larger values of $\mu_{\gamma \gamma}$, as it can be seen in the middle and bottom panels.

%

%%%%%%%%%%%%%%%%%%%%%%%%%%%%%%%%%%%%%%%%%%%%%%%%%%%%%%%%%%%%%%%%%%%%%%%%%%%%%
\section{Discussion}
\label{sec:discussion}
%%%%%%%%%%%%%%%%%%%%%%%%%%%%%%%%%%%%%%%%%%%%%%%%%%%%%%%%%%%%%%%%%%%%%%%%%%%%%

Excesses reported by CMS, ATLAS and LEP, in the ditau, diphoton and dibottom channels, are all compatible with the hypothesis that a new 95 GeV Higgs scalar exists. Although the overlap of these experimental results hint at a potential novel particle, the recent searches carried out by CMS and ATLAS for diphoton resonances were not able to confirm the 95~GeV excess, maintaining its local significance and diminishing the signal strength, from $\mu_{\gamma \gamma} = 0.6 \pm 0.2$ with 2016 CMS data~\cite{Biekotter:2022jyr} to $\mu_{\gamma \gamma} = 0.33^{+0.19}_{-0.12}$ with the full Run 2 data~\cite{Biekotter:2023jld}, to $\mu_{\gamma \gamma} = 0.27^{+0.10}_{-0.09}$ with ATLAS+CMS data~\cite{Biekotter:2023oen}. These results motivate a different approach than the one usually taken to look at the three 95~GeV experimental anomalies, where one or two of them are regarded as statistical fluctuations that should disappear once more data is collected. Here we follow this approach within the context of the UN2HDM, a framework motivated by an intriguing 2 TeV excess which can be addressed with a new vector boson. Besides featuring an extended gauge sector, which requires additional VLLs for anomaly cancellation, the scalar sector is comprised of a 2HDM with an extra complex scalar singlet. We construct the simplest possible type-I and III Yukawa sectors of the UN2HDM and analyse in detail their prospects in regards to the aforementioned excesses. 

None of the scenarios can accommodate the three excesses simultaneously at the 1$\sigma$ level. Hence, if all hints are confirmed in the future, the type-I and III UN2HDMs considered here with VLLs will be ruled out; these scenarios are therefore falsifiable. In the more likely case that one or two, or even all excesses disappear, we provide full insight on the capabilities of these two UN2HDM versions. Namely, these models can accommodate
\begin{itemize}
\item $b\bar b$ and $\gamma \gamma$ at the 1$\sigma$ level (both type-I and type-III);
\item $b\bar b$ and $\tau \tau$ at the 1$\sigma$ (type-III);
\item $b \bar b$ without excesses in the remaining channels (both type-I and type-III);
\item $\tau \tau$ without excesses in the remaining channels (type-III).
\end{itemize}
The different possibilities can be easily read from Figs.~\ref{fig:UN2HDMTImus} and \ref{fig:UN2HDMTIchi2}.

These distinct features of the two models can be understood from a theoretical viewpoint, looking at the couplings between the scalars and fermions in the Yukawa sector. Namely, since in type-I the same doublet couples to all fermions, it is not possible to accommodate simultaneously $b\bar{b}$ and $\tau \tau$ excesses. However, in type-III the doublets couple in distinct manner to the leptons and quarks allowing to explain either $\tau \tau$ or $b\bar{b}$. Nonetheless, experimental constraints on $\tan\beta$ favor higher values of this parameter, making a simultaneous explanation of both these excesses difficult. On the other hand, the coupling of all quarks to the same Higgs doublet in type-I and type-III UN2HDM explains the impossibility of accommodating the 95~GeV diphoton excess at $1\sigma$ level in both models. Despite the new loop contributions to the $H_1\rightarrow \gamma\gamma$ decay provided by the charged VLLs and the charged scalar, these are not significant enough to compensate for the lack of a mechanism in our models to enhance the Yukawa coupling to top quarks (which provides a dominant loop contribution to the decay and to Higgs production via gluon-gluon fusion) while suppressing the Yukawa coupling to bottom-quarks. This is due to the fact that the couplings between the scalar and charged VLLs are proportional $v/u$, the VEV $u$ value is set by the $Z^\prime$ mass to be around $2$ TeV (20 TeV) for type-I (type-III), suppressing this contribution [see Eqs.~\eqref{eq:typeIVLL} and~\eqref{eq:typeIIIVLL} of Appendix~\ref{sec:interactions}]. 

In summary, we confronted in a thorough analysis the recent experimental hints for a new 95 GeV Higgs boson, within the type-I and III UN2HDM, featuring VLLs. We took the point of view that the diminishing $\gamma \gamma$ excess indicates that these hints might as well disappear in the future with more experimental data. Therefore, we studied all possible scenarios where one or two of these excesses vanish, showing interesting connections between the signal strength predictions of the models. It is clear that in the very unlikely event that all excesses are confirmed these two versions of the UN2HDM are excluded. In such case, one could envisage to find the minimal realisations of the type-II and IV UN2HDM, which would require more exotic fermion content besides VLLs such as VLQs for anomaly cancellation. This is due to the fact that, for these Yukawa structures up and down quarks couple to different scalar doublets. Although it does not seem clear these excesses will prevail, studies such as the one presented here should be pursued with the goal of shedding light on possible new physics discoveries.

%%%%%%%%%%%%%%%%%%%%%%%%%%%%%%%%%%%%%%%%%%%%%%%%%%%%%%%%%%%%%%%%%%%%%%%%%%%%%
\section*{Acknowledgements}

We thank Thomas Biek\"otter and Fabio Campello for communications on several aspects related to \texttt{HiggsTools} and \texttt{EVADE}. The work of J.A.A.S. has been supported by MICINN project PID2019-110058GB-C21 and CEX2020-001007-S funded by MCIN/AEI/10.13039/501100011033 and by ERDF. This work is financially supported by Funda\c{c}{\~a}o para a Ci{\^e}ncia e a Tecnologia (FCT, Portugal) through the projects UIDB/00777/2020, UIDP/00777/2020 and CERN/FIS-PAR/0019/2021. The work of H.B.C. and J.F.S. is supported by the PhD FCT grants 2021.06340.BD and SFRH/BD/143891/2019, respectively. H.B.C.
thanks IFT/UAM-CSIC (Madrid) for hospitality and financial support
during the final stage of this work.
%%%%%%%%%%%%%%%%%%%%%%%%%%%%%%%%%%%%%%%%%%%%%%%%%%%%%%%%%%%%%%%%%%%%%%%%%%%%%

\appendix

%%%%%%%%%%%%%%%%%%%%%%%%%%%%%%%%%%%%%%%%%%%%%%%%%%%%%%%%%%%%%%%%%%%%%%%%%%%%%
\section{Parameter reconstruction}
\label{sec:parameters}
%%%%%%%%%%%%%%%%%%%%%%%%%%%%%%%%%%%%%%%%%%%%%%%%%%%%%%%%%%%%%%%%%%%%%%%%%%%%%

The scalar potential of the UN2HDM, presented in Sec.~\ref{sec:model}, has eleven parameters, which matches the number of physical parameters of the scalar sector of the theory. The VEV of the scalar singlet can be expressed as 
\begin{equation}
u^2 = \frac{m_Z^2 + m_{Z'}^2 - m_W^2/c_W^2}{(g_{Z'}Y')^2} - v^2\cbeta^2\,,
\end{equation}
in terms of the U($1)^\prime$ group coupling constant $g_{Z^\prime}$ and $Z^\prime$ gauge boson mass $m_{Z^\prime}$ (see Table~\ref{tab:TypeI}, \ref{tab:TypeIII} and \ref{tab:Scan}). In the above equation $m_{Z,W}$ are the usual $Z$ and $W$-boson masses and $c_W \equiv \cos \theta_W$, with $\theta_W$ being the weak angle.
Inverting Eq.~(\ref{eq:mA0}), we can write
\begin{equation}
\mu = -\frac{\sqrt{2}u\cbeta\sbeta}{u^2 + v^2\cbeta^2\sbeta^2}\,m_{A^0}^2\,,
\label{eq:mu}
\end{equation}
in terms of VEV parameters and the pseudoscalar mass $m_{A^0}$. The quartic coupling $\lambda_4$ is determined through the charged-scalar mass in Eq.~(\ref{eq:Mch}),
\begin{equation}
\lambda_4 = -\frac{1}{v^2}\left[2 m_{H^\pm}^2 + \frac{\sqrt{2}u\mu}{s_\beta c_\beta}\right]\,.
\label{eq:lambda 4}
\end{equation}
The remaining quartic couplings $\lambda_i$ can be expressed in terms of the above parameters, as well as the neutral scalar masses and mixings of Eqs.~(\ref{eq:O}) and~(\ref{eq:MH}). Defining $\tilde{M}^2=(m_h^2,m_{H_1}^2,m_{H_2}^2)$, we have
\begin{align}
\lambda_1 & = \frac{1}{v^2\cbeta^2}\left[\displaystyle\sum_{i=1}^3 \tilde{M}_i^2 O_{1i}^2 + \frac{u\mu}{\sqrt{2}}\frac{\sbeta}{\cbeta}\right] \; , \;
\lambda_2 = \frac{1}{v^2\sbeta^2}\left[\displaystyle\sum_{i=1}^3 \tilde{M}_i^2 O_{2i}^2 + \frac{u\mu}{\sqrt{2}}\frac{\cbeta}{\sbeta}\right] \; , \nonumber \\
\lambda_3 & = \frac{1}{v^2\cbeta\sbeta}\left[\displaystyle\sum_{i=1}^3 \tilde{M}_i^2 O_{1i}O_{2i} - \frac{u\mu}{\sqrt{2}}\right] - \lambda_4 \; , \;
\lambda_5 = \frac{1}{u^2}\left[\displaystyle\sum_{i=1}^3 \tilde{M}_i^2 O_{3i}^2 + \frac{v^2\mu}{\sqrt{2}u}\cbeta\sbeta\right] \; , \nonumber \\
\lambda_6 & = \frac{2}{uv\cbeta}\left[\displaystyle\sum_{i=1}^3 \tilde{M}_i^2 O_{1i}O_{3i} - \frac{v\mu}{\sqrt{2}}\sbeta\right] \; , \;
\lambda_7 = \frac{2}{uv\sbeta}\left[\displaystyle\sum_{i=1}^3 \tilde{M}_i^2 O_{2i}O_{3i} - \frac{v\mu}{\sqrt{2}}\cbeta\right] \,.
\label{eq:lambdas}
\end{align}
%

%%%%%%%%%%%%%%%%%%%%%%%%%%%%%%%%%%%%%%%%%%%%%%%%%%%%%%%%%%%%%%%%%%%%%%%%%%%%%
\section{Scalar couplings in the mass-eigenstate basis}
\label{sec:interactions}
%%%%%%%%%%%%%%%%%%%%%%%%%%%%%%%%%%%%%%%%%%%%%%%%%%%%%%%%%%%%%%%%%%%%%%%%%%%%%

Here we collect the interactions between the scalars $h/H_{1,2}/A^0$ and the $W$-boson, charged Higgs $H^{\pm}$ and fermions for both the type-I and III Yukawa sectors. These are the relevant interactions entering the Higgs to diphoton one-loop computation (see Appendix~\ref{sec:diphoton}). As for the remaining scalar vector-boson and triple scalar interactions, these can be found in Ref.~\cite{Aguilar-Saavedra:2022rvy}. Defining the CP-even neutral scalar vector as $\tilde{H} = (h, H_1, H_2)$, the Lagrangian terms involving two $W$-bosons and one scalar can be written as
\begin{equation}
\mathcal{L}_{\tilde{H}_i W^+W^-} = g_{\tilde{H}_i W^+W^-} {W^-}^\mu W^+_\mu \tilde{H}_i + {\rm H.c.} \; , \; g_{\tilde{H}_i W^+W^-} = \frac{g^2 v}{2}(O_{1i}\cbeta + O_{2i}\sbeta) \,,
\label{eq:SVV} 
\end{equation}
with the $\beta$ angle and CP-even scalar rotation matrix $O$ defined in Eqs.~\eqref{eq:VEV} and~\eqref{eq:O}, respectively. Notice that, since~$A^0$ is a pseudoscalar it does not interact with the $W$-boson.

The interactions of one neutral scalar and two charged ones are
\begin{equation}
\mathcal{L}_{\tilde{H}_iH^+H^-} =  v  \displaystyle g_{\tilde{H}_iH^+H^-} \tilde{H}_i H^+H^- \; , \; g_{\tilde{H}_iH^+H^-} = - \displaystyle\sum_{p=1}^3\sum_{q,r=1}^2 \,C^{\rm c}_{pqr} O_{pi}U_{q2}U_{r2}\,,
\label{eq: VertexHHchHch}
\end{equation}
where the charged scalar rotation matrix $U$ is defined in Eq.~(\ref{eq:U}). Note that, due to the pseudoscalar character of $A^0$, it does not interact with the charged Higgs $H^{\pm}$. The nonzero coefficients $C^{\rm c}_{pqr}$ are
\begin{align}
C^{\rm c}_{111} & = \lambda_1 \cbeta \; , \;
C^{\rm c}_{112} = C^{\rm c}_{121} = \frac{1}{2}\lambda_4 \sbeta \; , \;
C^{\rm c}_{122} = \lambda_3 \cbeta \; , \;
C^{\rm c}_{211} = \lambda_3 \sbeta \,, \notag \\
C^{\rm c}_{212} &= C^{\rm c}_{221} = \frac{1}{2}\lambda_4 \cbeta \; , \;
C^{\rm c}_{222} = \lambda_2 \sbeta \; , \;
C^{\rm c}_{311} = \frac{1}{2}\lambda_6 \frac{u}{v} \; , \;
C^{\rm c}_{312} = C^{\rm c}_{321} = \frac{1}{\sqrt{2}v}\mu \,, \notag \\
C^{\rm c}_{322} & = \frac{1}{2}\lambda_7 \frac{u}{v} \; , \; C
^{\rm c}_{412} = -C^{\rm c}_{421} = -\frac{i}{2}\lambda_4 \sbeta \; , \; 
C^{\rm c}_{512} = -C^{\rm c}_{521} = \frac{i}{2}\lambda_4 \cbeta \; , \; 
C^{\rm c}_{612} = -C^{\rm c}_{621} = -\frac{i}{\sqrt{2}v}\mu \,,
\end{align}
written in terms of the VEVs and scalar potential parameters of Eq.~\eqref{eq:V}.

The couplings among the scalars $S\equiv \tilde{H}_i, A^0$ ($i=1,2,3$) and fermions $f=e,d,u,E$, are generically written as
\begin{equation}
\mathcal{L}_{S f \overline{f}}= - \sum_f \frac{m_f}{v} \; \overline{f} \left( a_S^f + i b_S^f \gamma_5 \right) f S \; , 
\end{equation}
where $a_S^f$ and $b_S^f$ are the scalar-fermion and pseudoscalar-fermion couplings, respectively. 
For the type-I Yukawa model~(see Sec.~\ref{sec:TypeI}) we have for the couplings with SM fermions
\begin{align}
    a_{\text{I},\tilde{H}_i}^{e,d,u} = \frac{O_{2i}}{s_\beta} \; , \; b_{\text{I},A^0}^{e,d} = \frac{R_{23}}{s_\beta} \; , \; b_{\text{I},A^0}^{u} = - \frac{R_{23}}{s_\beta} \; ,
\end{align}
with the rotation matrix $R$ being defined in Eq.~\eqref{eq:Rrot}. For the couplings involving charged VLLs we have
\begin{align}
    a_{\text{I},\tilde{H}_i}^{E_j} = \frac{v}{m_{E_j}} C_{jj i}^E \; , \; 
    b_{\text{I},\tilde{H}_i}^{E_j} = \frac{v}{m_{E_j}} D_{jj i}^E \; , \;
     a_{\text{I},A^0}^{E_j} = \frac{v}{m_{E_j}} \tilde{C}_{jj}^E \; , \; 
    b_{\text{I},A^0}^{E_j} = \frac{v}{m_{E_j}} \tilde{D}_{jj}^E \; ,
    \label{eq:typeIVLL}
\end{align}
where,
\begin{align}
    C_{jkl}^E  = \frac{1}{2} \left(A_{jkl} + A_{kjl}^\ast \right) \; , \;
    D_{jkl}^E = -\frac{i}{2} \left(A_{jkl} - A_{kjl}^\ast \right) \; , \;
    \tilde{C}_{jk}^E = \frac{i}{2} \left(B_{jk} - B_{kj}^\ast \right) \; , \;
    \tilde{D}_{jk}^E = \frac{1}{2} \left(B_{jk} + B_{kj}^\ast \right) \; ,
\end{align}
with
\begin{align}
    A_{jkl}^E & = \frac{O_{2 l}}{\sqrt{2}}   \left[ w_1^E (\mathbf{V}_L^E)^{\ast}_{j 1} (\mathbf{V}_R^E)_{2 k} + w_2^E (\mathbf{V}_L^E)^{\ast}_{j 2} (\mathbf{V}_R^E)_{1 k}\; \right] + \frac{O_{3 l}}{\sqrt{2}}  \left[ y_2^E (\mathbf{V}_L^E)^{\ast}_{j 2} (\mathbf{V}_R^E)_{2 k} + y_1 (\mathbf{V}_L^E)^{\ast}_{j 1} (\mathbf{V}_R^E)_{1 k}\; \right] , \nonumber \\
   B_{jk}^E & = \frac{R_{2 3}}{\sqrt{2}}  \left[ w_1^E (\mathbf{V}_L^E)^{\ast}_{j 1} (\mathbf{V}_R^E)_{2 k} - w_2^E (\mathbf{V}_L^E)^{\ast}_{j 2} (\mathbf{V}_R^E)_{1 k}\; \right] + \frac{R_{3 3}}{\sqrt{2}}  \left[y_2^E (\mathbf{V}_L^E)^{\ast}_{j 2} (\mathbf{V}_R^E)_{2 k} - y_1 (\mathbf{V}_L^E)^{\ast}_{j 1} (\mathbf{V}_R^E)_{1 k}\; \right] ,
\end{align}
written in terms of the Yukawa couplings of Eq.~\eqref{eq:YukawasVLLI} and the $2 \times 2$ unitary matrix rotations of the charged VLLs defined in Eq.~\eqref{eq:VLRTI}. For the type-III model~(see Sec.~\ref{sec:TypeI}), the interaction couplings with SM fermions are given by
\begin{align}
    a_{\text{III},\tilde{H}_i}^{e} = \frac{O_{1i}}{c_\beta} \; , \; a_{\text{III},\tilde{H}_i}^{d,u} = \frac{O_{2i}}{s_\beta} \; , \;
    b_{\text{III},A^0}^{e} = \frac{R_{13}}{c_\beta} \; , \; 
     b_{\text{III},A^0}^{d} = \frac{R_{23}}{s_\beta} \; , \; b_{\text{III},A^0}^{u} = - \frac{R_{23}}{s_\beta} \; ,
\end{align}
while the ones involving charged VLLs are written as
\begin{align}
    a_{\text{III},\tilde{H}_i}^{E} = O_{3i} \frac{v}{u} \; , \; b_{\text{III},A^0}^{E} = - R_{33} \frac{v}{u} \; .
    \label{eq:typeIIIVLL}
\end{align}
%

%%%%%%%%%%%%%%%%%%%%%%%%%%%%%%%%%%%%%%%%%%%%%%%%%%%%%%%%%%%%%%%%%%%%%%%%%%%%%
\section{Higgs to diphoton decay rate}
\label{sec:diphoton}
%%%%%%%%%%%%%%%%%%%%%%%%%%%%%%%%%%%%%%%%%%%%%%%%%%%%%%%%%%%%%%%%%%%%%%%%%%%%%

%
\begin{figure}[!t]
   \centering
   \includegraphics[scale=1.2]{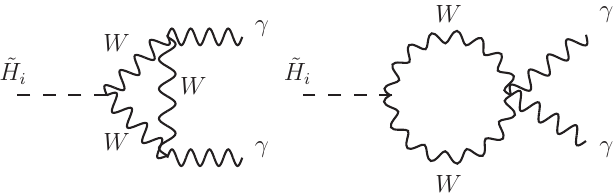} \\
   \includegraphics[scale=1.2]{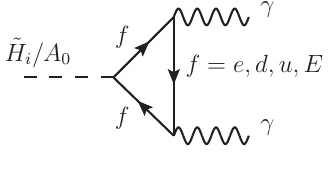} \hspace{+0.1cm} \includegraphics[scale=1.2]{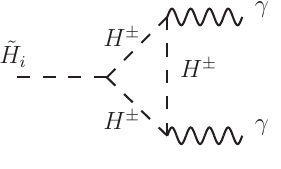}
  \caption{One-loop diagrams for $\tilde{H}_i/A^0 \rightarrow \gamma \gamma$ ($i=1,2,3$) in the UN2HDM, where the charged VLLs $E$ and Higgs $H^\pm$ are the BSM contributions.}
  \label{fig:diphoton}
\end{figure}

In Fig.~\ref{fig:diphoton} we present the one-loop diagrams contributing to the Higgs to diphoton decay rate $S \rightarrow \gamma \gamma$, in our framework, where $S\equiv \tilde{H}_i, A^0$ ($i=1,2,3$). Besides the usual SM loops ($W$-boson and SM fermions $e$, $d$ and $u$), the UN2HDM leads to BSM contributions to $S \rightarrow \gamma \gamma$ mediated by the charged VLLs $E$ and Higgs $H^{\pm}$. The decay width for this process is~\cite{Djouadi:2005gj,Posch:2010hx,Fontes:2014xva}
\begin{align}
\Gamma (S \rightarrow \gamma \gamma) = \frac{G_F \alpha^2 m_S^3}{128 \sqrt{2} \pi^3} \left(\left|C_W + C_f + C_{H^\pm}\right|^2 + |\widetilde{C}_f|^2 \right) \; ,
\end{align}
with the form factors $C_X$ ($X=W,f,H^{\pm}$) and $\widetilde{C}_f$ being
\begin{align}
C_W &= g_{SW^+W^-} \left[ 3 \tau_W + 2 \tau_W^2 + 3 \left(2 \tau_W -1\right) g(\tau_W) \right] \tau_W^{-2} \; ,\nonumber \\
C_f &= - 2 \sum_f N_c^f Q_f^2 a_S^f \left[\tau_f + \left(\tau_f - 1 \right) g(\tau_f) \right] \tau_f^{-2}\; , \nonumber \\
\widetilde{C}_f &= - 2 \sum_{f} N_c^f Q_f^2 b_S^f g(\tau_f) \tau_f^{-1} \; , \nonumber \\
C_{H^\pm} & = - \frac{v^2 g_{S H^+ H^{-}}}{2 m_{H^{\pm}}^2} \left[ \tau_{H^{\pm}} - g(\tau_{H^{\pm}})\right] \tau_{H^{\pm}}^{-2}\; ,
\end{align}
where $\tau_X=m_S^2/4 m_X^2$, $N_c^f$ and $Q_f$ are respectively the color factor and electric charge of fermions $f=e,d,u,E$. Furthermore, $g_{SW^+W^-}$, $a_S^f/b_S^f$ and $g_{S H^+ H^{-}}$ are the scalar coupling to the $W$-boson, fermions and charged Higgs $H^{\pm}$, respectively (see Appendix~\ref{sec:interactions}). Lastly, the loop function is given by
\begin{eqnarray}
  g(\tau) =\left\{  
  \begin{array}{ll}  \displaystyle
    \arcsin^2\sqrt{\tau} & \ \text{for} \ \tau\leq 1 \\
    \displaystyle -\frac{1}{4}\left[ \log\frac{1+\sqrt{1-\tau^{-1}}}
    {1-\sqrt{1-\tau^{-1}}}-i\pi \right]^2 \hspace{0.5cm} & \ \text{for} \ \tau>1
  \end{array} \right. .
\end{eqnarray}

%%%%%%%%%%%%%%%%%%%%%%%%%%%%%%%%%%%%%%%%%%%%%%%%%%%%%%%%%%%%%%%%%%%%%%%%%%%%%
\bibliographystyle{utphys}
\bibliography{bibliography}

\providecommand{\href}[2]{#2}\begingroup\raggedright\begin{thebibliography}{10}

\bibitem{CMS:2018cyk}
{\bfseries CMS} Collaboration, A.~M. Sirunyan {\em et~al.}, ``{Search for a
  standard model-like Higgs boson in the mass range between 70 and 110 GeV in
  the diphoton final state in proton-proton collisions at $\sqrt{s}=$ 8 and 13
  TeV},'' \href{http://dx.doi.org/10.1016/j.physletb.2019.03.064}{{\em Phys.
  Lett. B} {\bfseries 793} (2019) 320--347},
  \href{http://arxiv.org/abs/1811.08459}{{\ttfamily arXiv:1811.08459
  [hep-ex]}}.

\bibitem{ATLAS:2018xad}
{\bfseries ATLAS} Collaboration, ``{Search for resonances in the 65 to 110 GeV
  diphoton invariant mass range using 80 fb$^{-1}$ of $pp$ collisions collected
  at $\sqrt{s}=13$ TeV with the ATLAS detector},'' ATLAS-CONF-2018-025.
  \url{https://cds.cern.ch/record/2628760}.

\bibitem{CMS:2023yay}
{\bfseries CMS} Collaboration, ``{Search for a standard model-like Higgs boson
  in the mass range between 70 and 110$~\mathrm{GeV}$ in the diphoton final
  state in proton-proton collisions at $\sqrt{s}=13~\mathrm{TeV}$},''
  CMS-PAS-HIG-20-002. \url{https://cds.cern.ch/record/2852907}.

\bibitem{ATLAS-CONF-2023-035}
{\bfseries ATLAS} Collaboration, ``{Search for diphoton resonances in the 66 to
  110 GeV mass range using 140 fb$^{-1}$ of 13 TeV $pp$ collisions collected
  with the ATLAS detector},'' ATLAS-CONF-2023-035.
  \url{http://cds.cern.ch/record/2862024}.

\bibitem{CMS:2022goy}
{\bfseries CMS} Collaboration, ``{Searches for additional Higgs bosons and for
  vector leptoquarks in $\tau\tau$ final states in proton-proton collisions at
  $\sqrt{s}$ = 13 TeV},'' \href{http://arxiv.org/abs/2208.02717}{{\ttfamily
  arXiv:2208.02717 [hep-ex]}}.

\bibitem{LEPWorkingGroupforHiggsbosonsearches:2003ing}
{\bfseries LEP Working Group for Higgs boson searches, ALEPH, DELPHI, L3, OPAL}
  Collaboration, R.~Barate {\em et~al.}, ``{Search for the standard model Higgs
  boson at LEP},'' \href{http://dx.doi.org/10.1016/S0370-2693(03)00614-2}{{\em
  Phys. Lett. B} {\bfseries 565} (2003) 61--75},
  \href{http://arxiv.org/abs/hep-ex/0306033}{{\ttfamily arXiv:hep-ex/0306033}}.

\bibitem{Azatov:2012bz}
A.~Azatov, R.~Contino, and J.~Galloway, ``{Model-Independent Bounds on a Light
  Higgs},'' \href{http://dx.doi.org/10.1007/JHEP04(2012)127}{{\em JHEP}
  {\bfseries 04} (2012) 127}, \href{http://arxiv.org/abs/1202.3415}{{\ttfamily
  arXiv:1202.3415 [hep-ph]}}. [Erratum: JHEP 04, 140 (2013)].

\bibitem{Cao:2016uwt}
J.~Cao, X.~Guo, Y.~He, P.~Wu, and Y.~Zhang, ``{Diphoton signal of the light
  Higgs boson in natural NMSSM},''
  \href{http://dx.doi.org/10.1103/PhysRevD.95.116001}{{\em Phys. Rev. D}
  {\bfseries 95} no.~11, (2017) 116001},
  \href{http://arxiv.org/abs/1612.08522}{{\ttfamily arXiv:1612.08522
  [hep-ph]}}.

\bibitem{Biekotter:2023oen}
T.~Biek\"otter, S.~Heinemeyer, and G.~Weiglein, ``{The 95.4 GeV di-photon
  excess at ATLAS and CMS},'' \href{http://arxiv.org/abs/2306.03889}{{\ttfamily
  arXiv:2306.03889 [hep-ph]}}.

\bibitem{Fox:2017uwr}
P.~J. Fox and N.~Weiner, ``{Light Signals from a Lighter Higgs},''
  \href{http://dx.doi.org/10.1007/JHEP08(2018)025}{{\em JHEP} {\bfseries 08}
  (2018) 025}, \href{http://arxiv.org/abs/1710.07649}{{\ttfamily
  arXiv:1710.07649 [hep-ph]}}.

\bibitem{Haisch:2017gql}
U.~Haisch and A.~Malinauskas, ``{Let there be light from a second light Higgs
  doublet},'' \href{http://dx.doi.org/10.1007/JHEP03(2018)135}{{\em JHEP}
  {\bfseries 03} (2018) 135}, \href{http://arxiv.org/abs/1712.06599}{{\ttfamily
  arXiv:1712.06599 [hep-ph]}}.

\bibitem{Azevedo:2023zkg}
D.~Azevedo, T.~Biek\"otter, and P.~M. Ferreira, ``{2HDM interpretations of the
  CMS diphoton excess at 95 GeV},''
  \href{http://arxiv.org/abs/2305.19716}{{\ttfamily arXiv:2305.19716
  [hep-ph]}}.

\bibitem{Belyaev:2023xnv}
A.~Belyaev, R.~Benbrik, M.~Boukidi, M.~Chakraborti, S.~Moretti, and S.~Semlali,
  ``{Explanation of the Hints for a 95 GeV Higgs Boson within a 2-Higgs Doublet
  Model},'' \href{http://arxiv.org/abs/2306.09029}{{\ttfamily arXiv:2306.09029
  [hep-ph]}}.

\bibitem{Biekotter:2019kde}
T.~Biek\"otter, M.~Chakraborti, and S.~Heinemeyer, ``{A 96 GeV Higgs boson in
  the N2HDM},'' \href{http://dx.doi.org/10.1140/epjc/s10052-019-7561-2}{{\em
  Eur. Phys. J. C} {\bfseries 80} no.~1, (2020) 2},
  \href{http://arxiv.org/abs/1903.11661}{{\ttfamily arXiv:1903.11661
  [hep-ph]}}.

\bibitem{Biekotter:2021qbc}
T.~Biek\"otter, A.~Grohsjean, S.~Heinemeyer, C.~Schwanenberger, and
  G.~Weiglein, ``{Possible indications for new Higgs bosons in the reach of the
  LHC: N2HDM and NMSSM interpretations},''
  \href{http://dx.doi.org/10.1140/epjc/s10052-022-10099-1}{{\em Eur. Phys. J.
  C} {\bfseries 82} no.~2, (2022) 178},
  \href{http://arxiv.org/abs/2109.01128}{{\ttfamily arXiv:2109.01128
  [hep-ph]}}.

\bibitem{Heinemeyer:2021msz}
S.~Heinemeyer, C.~Li, F.~Lika, G.~Moortgat-Pick, and S.~Paasch,
  ``{Phenomenology of a 96~GeV Higgs boson in the 2HDM with an additional
  singlet},'' \href{http://dx.doi.org/10.1103/PhysRevD.106.075003}{{\em Phys.
  Rev. D} {\bfseries 106} no.~7, (2022) 075003},
  \href{http://arxiv.org/abs/2112.11958}{{\ttfamily arXiv:2112.11958
  [hep-ph]}}.

\bibitem{Biekotter:2022jyr}
T.~Biek\"otter, S.~Heinemeyer, and G.~Weiglein, ``{Mounting evidence for a 95
  GeV Higgs boson},'' \href{http://dx.doi.org/10.1007/JHEP08(2022)201}{{\em
  JHEP} {\bfseries 08} (2022) 201},
  \href{http://arxiv.org/abs/2203.13180}{{\ttfamily arXiv:2203.13180
  [hep-ph]}}.

\bibitem{Biekotter:2021ovi}
T.~Biek\"otter and M.~O. Olea-Romacho, ``{Reconciling Higgs physics and
  pseudo-Nambu-Goldstone dark matter in the S2HDM using a genetic algorithm},''
  \href{http://dx.doi.org/10.1007/JHEP10(2021)215}{{\em JHEP} {\bfseries 10}
  (2021) 215}, \href{http://arxiv.org/abs/2108.10864}{{\ttfamily
  arXiv:2108.10864 [hep-ph]}}.

\bibitem{Biekotter:2023jld}
T.~Biek\"otter, S.~Heinemeyer, and G.~Weiglein, ``{The CMS di-photon excess at
  95 GeV in view of the LHC Run 2 results},''
  \href{http://arxiv.org/abs/2303.12018}{{\ttfamily arXiv:2303.12018
  [hep-ph]}}.

\bibitem{Aguilar-Saavedra:2020wrj}
J.~A. Aguilar-Saavedra and F.~R. Joaquim, ``{Multiphoton signals of a (96 GeV?)
  stealth boson},''
  \href{http://dx.doi.org/10.1140/epjc/s10052-020-7952-4}{{\em Eur. Phys. J. C}
  {\bfseries 80} no.~5, (2020) 403},
  \href{http://arxiv.org/abs/2002.07697}{{\ttfamily arXiv:2002.07697
  [hep-ph]}}.

\bibitem{Ashanujjaman:2023etj}
S.~Ashanujjaman, S.~Banik, G.~Coloretti, A.~Crivellin, B.~Mellado, and A.-T.
  Mulaudzi, ``{$SU(2)_L$ triplet scalar as the origin of the 95 GeV excess?},''
  \href{http://arxiv.org/abs/2306.15722}{{\ttfamily arXiv:2306.15722
  [hep-ph]}}.

\bibitem{Banik:2023ecr}
S.~Banik, A.~Crivellin, S.~Iguro, and T.~Kitahara, ``{Asymmetric Di-Higgs
  Signals of the N2HDM-$U(1)$},''
  \href{http://arxiv.org/abs/2303.11351}{{\ttfamily arXiv:2303.11351
  [hep-ph]}}.

\bibitem{Biekotter:2017xmf}
T.~Biek\"otter, S.~Heinemeyer, and C.~Mu\~noz, ``{Precise prediction for the
  Higgs-boson masses in the $\mu \nu $ SSM},''
  \href{http://dx.doi.org/10.1140/epjc/s10052-018-5978-7}{{\em Eur. Phys. J. C}
  {\bfseries 78} no.~6, (2018) 504},
  \href{http://arxiv.org/abs/1712.07475}{{\ttfamily arXiv:1712.07475
  [hep-ph]}}.

\bibitem{Domingo:2018uim}
F.~Domingo, S.~Heinemeyer, S.~Pa\ss{}ehr, and G.~Weiglein, ``{Decays of the
  neutral Higgs bosons into SM fermions and gauge bosons in the
  $\mathcal{CP}$-violating NMSSM},''
  \href{http://dx.doi.org/10.1140/epjc/s10052-018-6400-1}{{\em Eur. Phys. J. C}
  {\bfseries 78} no.~11, (2018) 942},
  \href{http://arxiv.org/abs/1807.06322}{{\ttfamily arXiv:1807.06322
  [hep-ph]}}.

\bibitem{Cao:2019ofo}
J.~Cao, X.~Jia, Y.~Yue, H.~Zhou, and P.~Zhu, ``{96 GeV diphoton excess in
  seesaw extensions of the natural NMSSM},''
  \href{http://dx.doi.org/10.1103/PhysRevD.101.055008}{{\em Phys. Rev. D}
  {\bfseries 101} no.~5, (2020) 055008},
  \href{http://arxiv.org/abs/1908.07206}{{\ttfamily arXiv:1908.07206
  [hep-ph]}}.

\bibitem{Li:2022etb}
W.~Li, J.~Zhu, K.~Wang, S.~Ma, P.~Tian, and H.~Qiao, ``{A light Higgs boson in
  the NMSSM confronted with the CMS di-photon and di-tau excesses},''
  \href{http://arxiv.org/abs/2212.11739}{{\ttfamily arXiv:2212.11739
  [hep-ph]}}.

\bibitem{Richard:2017kot}
F.~Richard, ``{Search for a light radion at HL-LHC and ILC250},''
  \href{http://arxiv.org/abs/1712.06410}{{\ttfamily arXiv:1712.06410
  [hep-ex]}}.

\bibitem{Liu:2018xsw}
D.~Liu, J.~Liu, C.~E.~M. Wagner, and X.-P. Wang, ``{A Light Higgs at the LHC
  and the B-Anomalies},'' \href{http://dx.doi.org/10.1007/JHEP06(2018)150}{{\em
  JHEP} {\bfseries 06} (2018) 150},
  \href{http://arxiv.org/abs/1805.01476}{{\ttfamily arXiv:1805.01476
  [hep-ph]}}.

\bibitem{Cline:2019okt}
J.~M. Cline and T.~Toma, ``{Pseudo-Goldstone dark matter confronts cosmic ray
  and collider anomalies},''
  \href{http://dx.doi.org/10.1103/PhysRevD.100.035023}{{\em Phys. Rev. D}
  {\bfseries 100} no.~3, (2019) 035023},
  \href{http://arxiv.org/abs/1906.02175}{{\ttfamily arXiv:1906.02175
  [hep-ph]}}.

\bibitem{Escribano:2023hxj}
P.~Escribano, V.~M. Lozano, and A.~Vicente, ``{A Scotogenic explanation for the
  95 GeV excesses},'' \href{http://arxiv.org/abs/2306.03735}{{\ttfamily
  arXiv:2306.03735 [hep-ph]}}.

\bibitem{ATLAS:2015xom}
{\bfseries ATLAS} Collaboration, G.~Aad {\em et~al.}, ``{Search for high-mass
  diboson resonances with boson-tagged jets in proton-proton collisions at $
  \sqrt{s}=8 $ TeV with the ATLAS detector},''
  \href{http://dx.doi.org/10.1007/JHEP12(2015)055}{{\em JHEP} {\bfseries 12}
  (2015) 055}, \href{http://arxiv.org/abs/1506.00962}{{\ttfamily
  arXiv:1506.00962 [hep-ex]}}.

\bibitem{Aguilar-Saavedra:2015rna}
J.~A. Aguilar-Saavedra, ``{Triboson interpretations of the ATLAS diboson
  excess},'' \href{http://dx.doi.org/10.1007/JHEP10(2015)099}{{\em JHEP}
  {\bfseries 10} (2015) 099}, \href{http://arxiv.org/abs/1506.06739}{{\ttfamily
  arXiv:1506.06739 [hep-ph]}}.

\bibitem{Aguilar-Saavedra:2015iew}
J.~A. Aguilar-Saavedra and F.~R. Joaquim, ``{Multiboson production in
  $W^{\prime}$ decays},'' \href{http://dx.doi.org/10.1007/JHEP01(2016)183}{{\em
  JHEP} {\bfseries 01} (2016) 183},
  \href{http://arxiv.org/abs/1512.00396}{{\ttfamily arXiv:1512.00396
  [hep-ph]}}.

\bibitem{ATLAS:2017zuf}
{\bfseries ATLAS} Collaboration, M.~Aaboud {\em et~al.}, ``{Search for diboson
  resonances with boson-tagged jets in $pp$ collisions at $\sqrt{s}=13$ TeV
  with the ATLAS detector},''
  \href{http://dx.doi.org/10.1016/j.physletb.2017.12.011}{{\em Phys. Lett. B}
  {\bfseries 777} (2018) 91--113},
  \href{http://arxiv.org/abs/1708.04445}{{\ttfamily arXiv:1708.04445
  [hep-ex]}}.

\bibitem{CMS:2017fgc}
{\bfseries CMS} Collaboration, A.~M. Sirunyan {\em et~al.}, ``{Search for
  massive resonances decaying into $WW$, $WZ$, $ZZ$, $qW$, and $qZ$ with dijet
  final states at $\sqrt{s}=13\text{ }\text{ }\mathrm{TeV}$},''
  \href{http://dx.doi.org/10.1103/PhysRevD.97.072006}{{\em Phys. Rev. D}
  {\bfseries 97} no.~7, (2018) 072006},
  \href{http://arxiv.org/abs/1708.05379}{{\ttfamily arXiv:1708.05379
  [hep-ex]}}.

\bibitem{Aguilar-Saavedra:2017zuc}
J.~A. Aguilar-Saavedra, ``{Stealth multiboson signals},''
  \href{http://dx.doi.org/10.1140/epjc/s10052-017-5289-4}{{\em Eur. Phys. J. C}
  {\bfseries 77} no.~10, (2017) 703},
  \href{http://arxiv.org/abs/1705.07885}{{\ttfamily arXiv:1705.07885
  [hep-ph]}}.

\bibitem{arXiv:2210.00043}
{\bfseries CMS} Collaboration, ``{Search for new heavy resonances decaying to
  WW, WZ, ZZ, WH, or ZH boson pairs in the all-jets final state in
  proton-proton collisions at $ \sqrt{s}= $ 13 TeV},'' tech. rep., CERN,
  Geneva, 2022.
\newblock \href{http://arxiv.org/abs/2210.00043}{{\ttfamily arXiv:2210.00043}}.
\newblock \url{https://cds.cern.ch/record/2835154}.

\bibitem{Aguilar-Saavedra:2019adu}
J.~A. Aguilar-Saavedra and F.~R. Joaquim, ``{The minimal stealth boson: models
  and benchmarks},'' \href{http://dx.doi.org/10.1007/JHEP10(2019)237}{{\em
  JHEP} {\bfseries 10} (2019) 237},
  \href{http://arxiv.org/abs/1905.12651}{{\ttfamily arXiv:1905.12651
  [hep-ph]}}.

\bibitem{Aguilar-Saavedra:2022rvy}
J.~A. Aguilar-Saavedra, F.~R. Joaquim, and J.~F. Seabra, ``{Multiboson signals
  in the UN2HDM},''
  \href{http://dx.doi.org/10.1140/epjc/s10052-022-11046-w}{{\em Eur. Phys. J.
  C} {\bfseries 82} no.~11, (2022) 1080},
  \href{http://arxiv.org/abs/2206.01200}{{\ttfamily arXiv:2206.01200
  [hep-ph]}}.

\bibitem{Branco:2011iw}
G.~C. Branco, P.~M. Ferreira, L.~Lavoura, M.~N. Rebelo, M.~Sher, and J.~P.
  Silva, ``{Theory and phenomenology of two-Higgs-doublet models},''
  \href{http://dx.doi.org/10.1016/j.physrep.2012.02.002}{{\em Phys. Rept.}
  {\bfseries 516} (2012) 1--102},
  \href{http://arxiv.org/abs/1106.0034}{{\ttfamily arXiv:1106.0034 [hep-ph]}}.

\bibitem{ParticleDataGroup:2022pth}
{\bfseries Particle Data Group} Collaboration, R.~L. Workman {\em et~al.},
  ``{Review of Particle Physics},''
  \href{http://dx.doi.org/10.1093/ptep/ptac097}{{\em PTEP} {\bfseries 2022}
  (2022) 083C01}.

\bibitem{Erler:2009jh}
J.~Erler, P.~Langacker, S.~Munir, and E.~Rojas, ``{Improved Constraints on
  Z-prime Bosons from Electroweak Precision Data},''
  \href{http://dx.doi.org/10.1088/1126-6708/2009/08/017}{{\em JHEP} {\bfseries
  08} (2009) 017}, \href{http://arxiv.org/abs/0906.2435}{{\ttfamily
  arXiv:0906.2435 [hep-ph]}}.

\bibitem{Muhlleitner:2020wwk}
M.~M\"uhlleitner, M.~O.~P. Sampaio, R.~Santos, and J.~Wittbrodt, ``{ScannerS:
  parameter scans in extended scalar sectors},''
  \href{http://dx.doi.org/10.1140/epjc/s10052-022-10139-w}{{\em Eur. Phys. J.
  C} {\bfseries 82} no.~3, (2022) 198},
  \href{http://arxiv.org/abs/2007.02985}{{\ttfamily arXiv:2007.02985
  [hep-ph]}}.

\bibitem{Muhlleitner:2016mzt}
M.~Muhlleitner, M.~O.~P. Sampaio, R.~Santos, and J.~Wittbrodt, ``{The N2HDM
  under Theoretical and Experimental Scrutiny},''
  \href{http://dx.doi.org/10.1007/JHEP03(2017)094}{{\em JHEP} {\bfseries 03}
  (2017) 094}, \href{http://arxiv.org/abs/1612.01309}{{\ttfamily
  arXiv:1612.01309 [hep-ph]}}.

\bibitem{Hollik:2018wrr}
W.~G. Hollik, G.~Weiglein, and J.~Wittbrodt, ``{Impact of Vacuum Stability
  Constraints on the Phenomenology of Supersymmetric Models},''
  \href{http://dx.doi.org/10.1007/JHEP03(2019)109}{{\em JHEP} {\bfseries 03}
  (2019) 109}, \href{http://arxiv.org/abs/1812.04644}{{\ttfamily
  arXiv:1812.04644 [hep-ph]}}.

\bibitem{Ferreira:2019iqb}
P.~M. Ferreira, M.~M\"uhlleitner, R.~Santos, G.~Weiglein, and J.~Wittbrodt,
  ``{Vacuum Instabilities in the N2HDM},''
  \href{http://dx.doi.org/10.1007/JHEP09(2019)006}{{\em JHEP} {\bfseries 09}
  (2019) 006}, \href{http://arxiv.org/abs/1905.10234}{{\ttfamily
  arXiv:1905.10234 [hep-ph]}}.

\bibitem{Grimus:2007if}
W.~Grimus, L.~Lavoura, O.~M. Ogreid, and P.~Osland, ``{A Precision constraint
  on multi-Higgs-doublet models},''
  \href{http://dx.doi.org/10.1088/0954-3899/35/7/075001}{{\em J. Phys. G}
  {\bfseries 35} (2008) 075001},
  \href{http://arxiv.org/abs/0711.4022}{{\ttfamily arXiv:0711.4022 [hep-ph]}}.

\bibitem{Grimus:2008nb}
W.~Grimus, L.~Lavoura, O.~M. Ogreid, and P.~Osland, ``{The Oblique parameters
  in multi-Higgs-doublet models},''
  \href{http://dx.doi.org/10.1016/j.nuclphysb.2008.04.019}{{\em Nucl. Phys. B}
  {\bfseries 801} (2008) 81--96},
  \href{http://arxiv.org/abs/0802.4353}{{\ttfamily arXiv:0802.4353 [hep-ph]}}.

\bibitem{Haller:2018nnx}
J.~Haller, A.~Hoecker, R.~Kogler, K.~M\"onig, T.~Peiffer, and J.~Stelzer,
  ``{Update of the global electroweak fit and constraints on two-Higgs-doublet
  models},'' \href{http://dx.doi.org/10.1140/epjc/s10052-018-6131-3}{{\em Eur.
  Phys. J. C} {\bfseries 78} no.~8, (2018) 675},
  \href{http://arxiv.org/abs/1803.01853}{{\ttfamily arXiv:1803.01853
  [hep-ph]}}.

\bibitem{Engeln:2018mbg}
I.~Engeln, M.~M\"uhlleitner, and J.~Wittbrodt, ``{N2HDECAY: Higgs Boson Decays
  in the Different Phases of the N2HDM},''
  \href{http://dx.doi.org/10.1016/j.cpc.2018.07.020}{{\em Comput. Phys.
  Commun.} {\bfseries 234} (2019) 256--262},
  \href{http://arxiv.org/abs/1805.00966}{{\ttfamily arXiv:1805.00966
  [hep-ph]}}.

\bibitem{Bahl:2022igd}
H.~Bahl, T.~Biek\"otter, S.~Heinemeyer, C.~Li, S.~Paasch, G.~Weiglein, and
  J.~Wittbrodt, ``{HiggsTools: BSM scalar phenomenology with new versions of
  HiggsBounds and HiggsSignals},''
  \href{http://arxiv.org/abs/2210.09332}{{\ttfamily arXiv:2210.09332
  [hep-ph]}}.

\bibitem{Bechtle:2008jh}
P.~Bechtle, O.~Brein, S.~Heinemeyer, G.~Weiglein, and K.~E. Williams,
  ``{HiggsBounds: Confronting Arbitrary Higgs Sectors with Exclusion Bounds
  from LEP and the Tevatron},''
  \href{http://dx.doi.org/10.1016/j.cpc.2009.09.003}{{\em Comput. Phys.
  Commun.} {\bfseries 181} (2010) 138--167},
  \href{http://arxiv.org/abs/0811.4169}{{\ttfamily arXiv:0811.4169 [hep-ph]}}.

\bibitem{Bechtle:2011sb}
P.~Bechtle, O.~Brein, S.~Heinemeyer, G.~Weiglein, and K.~E. Williams,
  ``{HiggsBounds 2.0.0: Confronting Neutral and Charged Higgs Sector
  Predictions with Exclusion Bounds from LEP and the Tevatron},''
  \href{http://dx.doi.org/10.1016/j.cpc.2011.07.015}{{\em Comput. Phys.
  Commun.} {\bfseries 182} (2011) 2605--2631},
  \href{http://arxiv.org/abs/1102.1898}{{\ttfamily arXiv:1102.1898 [hep-ph]}}.

\bibitem{Bechtle:2013wla}
P.~Bechtle, O.~Brein, S.~Heinemeyer, O.~St\r{a}l, T.~Stefaniak, G.~Weiglein,
  and K.~E. Williams, ``{$\mathsf{HiggsBounds}-4$: Improved Tests of Extended
  Higgs Sectors against Exclusion Bounds from LEP, the Tevatron and the LHC},''
  \href{http://dx.doi.org/10.1140/epjc/s10052-013-2693-2}{{\em Eur. Phys. J. C}
  {\bfseries 74} no.~3, (2014) 2693},
  \href{http://arxiv.org/abs/1311.0055}{{\ttfamily arXiv:1311.0055 [hep-ph]}}.

\bibitem{Bechtle:2015pma}
P.~Bechtle, S.~Heinemeyer, O.~Stal, T.~Stefaniak, and G.~Weiglein, ``{Applying
  Exclusion Likelihoods from LHC Searches to Extended Higgs Sectors},''
  \href{http://dx.doi.org/10.1140/epjc/s10052-015-3650-z}{{\em Eur. Phys. J. C}
  {\bfseries 75} no.~9, (2015) 421},
  \href{http://arxiv.org/abs/1507.06706}{{\ttfamily arXiv:1507.06706
  [hep-ph]}}.

\bibitem{Bechtle:2020pkv}
P.~Bechtle, D.~Dercks, S.~Heinemeyer, T.~Klingl, T.~Stefaniak, G.~Weiglein, and
  J.~Wittbrodt, ``{HiggsBounds-5: Testing Higgs Sectors in the LHC 13 TeV
  Era},'' \href{http://dx.doi.org/10.1140/epjc/s10052-020-08557-9}{{\em Eur.
  Phys. J. C} {\bfseries 80} no.~12, (2020) 1211},
  \href{http://arxiv.org/abs/2006.06007}{{\ttfamily arXiv:2006.06007
  [hep-ph]}}.

\bibitem{Bechtle:2013xfa}
P.~Bechtle, S.~Heinemeyer, O.~St\r{a}l, T.~Stefaniak, and G.~Weiglein,
  ``{$HiggsSignals$: Confronting arbitrary Higgs sectors with measurements at
  the Tevatron and the LHC},''
  \href{http://dx.doi.org/10.1140/epjc/s10052-013-2711-4}{{\em Eur. Phys. J. C}
  {\bfseries 74} no.~2, (2014) 2711},
  \href{http://arxiv.org/abs/1305.1933}{{\ttfamily arXiv:1305.1933 [hep-ph]}}.

\bibitem{Bechtle:2020uwn}
P.~Bechtle, S.~Heinemeyer, T.~Klingl, T.~Stefaniak, G.~Weiglein, and
  J.~Wittbrodt, ``{HiggsSignals-2: Probing new physics with precision Higgs
  measurements in the LHC 13 TeV era},''
  \href{http://dx.doi.org/10.1140/epjc/s10052-021-08942-y}{{\em Eur. Phys. J.
  C} {\bfseries 81} no.~2, (2021) 145},
  \href{http://arxiv.org/abs/2012.09197}{{\ttfamily arXiv:2012.09197
  [hep-ph]}}.

\bibitem{CMS-PAS-EXO-21-018}
{\bfseries CMS} Collaboration, ``{Search for dilepton resonances from decays of
  (pseudo)scalar bosons produced in association with a massive vector boson or
  top quark anti-top quark pair at $\sqrt{s}=13~\mathrm{TeV}$},''
  CMS-PAS-EXO-21-018. \url{https://cds.cern.ch/record/2815307}.

\bibitem{Djouadi:2005gj}
A.~Djouadi, ``{The Anatomy of electro-weak symmetry breaking. II. The Higgs
  bosons in the minimal supersymmetric model},''
  \href{http://dx.doi.org/10.1016/j.physrep.2007.10.005}{{\em Phys. Rept.}
  {\bfseries 459} (2008) 1--241},
  \href{http://arxiv.org/abs/hep-ph/0503173}{{\ttfamily arXiv:hep-ph/0503173}}.

\bibitem{Posch:2010hx}
P.~Posch, ``{Enhancement of $h \to \gamma \gamma$ in the Two Higgs Doublet
  Model Type I},'' \href{http://dx.doi.org/10.1016/j.physletb.2011.01.003}{{\em
  Phys. Lett. B} {\bfseries 696} (2011) 447--453},
  \href{http://arxiv.org/abs/1001.1759}{{\ttfamily arXiv:1001.1759 [hep-ph]}}.

\bibitem{Fontes:2014xva}
D.~Fontes, J.~C. Rom\~ao, and J.~a.~P. Silva, ``{$h \rightarrow Z \gamma$ in
  the complex two Higgs doublet model},''
  \href{http://dx.doi.org/10.1007/JHEP12(2014)043}{{\em JHEP} {\bfseries 12}
  (2014) 043}, \href{http://arxiv.org/abs/1408.2534}{{\ttfamily arXiv:1408.2534
  [hep-ph]}}.

\end{thebibliography}\endgroup
%%%%%%%%%%%%%%%%%%%%%%%%%%%%%%%%%%%%%%%%%%%%%%%%%%%%%%%%%%%%%%%%%%%%%%%%%%%%%
%%%%% ScannerS

\end{document}